\documentclass{aa}  
\usepackage[normalem]{ulem}
\usepackage{gensymb}
\usepackage{amssymb}
\usepackage{graphicx}
\usepackage{txfonts}
\usepackage{lipsum}
\usepackage{xcolor}
\usepackage{amsmath}
\usepackage{multirow}
\usepackage[shortlabels]{enumitem}
\usepackage[section,above,below]{placeins}
\usepackage{subcaption} 
\usepackage[]{hyperref}
\begin{document} 

	\title{PDRs4All XVII\\ Formation and excitation of $\mathrm{HD}$ in photodissociation regions}
    \subtitle{Application to the Orion Bar}

   \author{
          Marion Zannese\inst{\ref{ias}}
          \and
          Jacques Le Bourlot\inst{\ref{lux},\ref{upc}}
          \and
          Evelyne Roueff\inst{\ref{lux}}
          \and Emeric Bron\inst{\ref{lux}} \and Franck Le Petit\inst{\ref{lux}} \and
          Dries Van De Putte \inst{\ref{uwo},\ref{iese}} \and Maryvonne Gerin\inst{\ref{lux}} \and Naslim Neelamkodan\inst{\ref{arab_emirates}} \and Javier R. Goicoechea\inst{\ref{iff}} \and John Black \inst{\ref{chalmers}} 
          \and 
          Ryan Chown \inst{\ref{uwo},\ref{iese}} \and
          Ameek Sidhu \inst{\ref{uwo},\ref{iese}} 
          \and
          Emilie Habart\inst{\ref{ias}} 
          \and
          Els Peeters \inst{\ref{uwo},\ref{iese}} 
          \and
          Olivier Bern\'{e} \inst{\ref{irap}} 
         }

   \institute{
    Institut d'Astrophysique Spatiale, Universit\'e Paris-Saclay, CNRS,  B$\hat{a}$timent 121, 91405 Orsay Cedex, France \label{ias}\\
    \email{marion.zannese@universite-paris-saclay.fr}
    \and
    LUX, Observatoire de Paris, Universit\'e PSL, Sorbonne Universit\'e, CNRS, 92190 Meudon, France \label{lux}
    \and
    Universit\'e Paris-Cit\'e, Paris, France \label{upc}
    \and
    Department of Physics \& Astronomy, The University of Western Ontario, London ON N6A 3K7, Canada; 
    \label{uwo}
    \and
    Institute for Earth and Space Exploration, The University of Western Ontario, London ON N6A 3K7, Canada \label{iese}
    \and Department of Physics, College of Science, United Arab Emirates University (UAEU), Al-Ain, 15551, USA \label{arab_emirates} \and Instituto de Física Fundamental (CSIC), Calle Serrano 121-123, 28006, Madrid, Spain \label{iff} \and
    Department of Space, Earth and Environment, Chalmers University of Technology, Onsala Space Observatory, 43992 Onsala, Sweden \label{chalmers}
    \and
    Institut de Recherche en Astrophysique et Plan\'etologie, Universit\'e Toulouse III - Paul Sabatier, CNRS, CNES, 9 Av. du colonel Roche, 31028 Toulouse Cedex 04, France \label{irap}
    }

   \date{Received; accepted }

  \abstract
  {The James Webb Space Telescope (JWST), with its high spatial resolution and sensitivity, enabled the first detection of several ro-vibrational emission lines of $\mathrm{HD}$ in the Orion Bar, a prototypical photodissociation region (PDR).
  This provides an incentive to 
  examine the physics of $\mathrm{HD}$ in dense and strong PDRs.
  }
   {Using the latest data available on $\mathrm{HD}$ excitation by collisional, radiative and chemical processes, our goal is to unveil $\mathrm{HD}$ formation and excitation processes in PDRs by comparing  our state-of-the-art PDR model with observations made in the Orion Bar and discuss if and how $\mathrm{HD}$ can be used as a complementary tracer of physical parameters (thermal pressure and intensity of the UV field) in the emitting region.}
   {
   We compute detailed PDR models, using an upgraded version of the Meudon PDR code (including $\mathrm{HD}$ radiative and collisional excitation of rovibrational levels).
   Model results are then compared to spectro-imaging data acquired using the NIRSpec instrument on board JWST using excitation diagrams and synthetic emission spectra.
   }
   {The models predict that $\mathrm{HD}$ is mainly produced in the gas phase via the reaction $\mathrm{D} + \mathrm{H_2} \rightarrow \mathrm{H} + \mathrm{HD}$ at the front edge of the PDR and that the $\mathrm{D/HD}$ transition is located slightly closer to the edge than the $\mathrm{H/H_2}$ transition. Rovibrational levels are excited by UV pumping. In the observations, 
   $\mathrm{HD}$ rovibrational emission is detected close to the $\mathrm{H}$/$\mathrm{H_2}$ dissociation fronts of the Orion Bar and peaks where vibrationally excited $\mathrm{H_2}$ peaks, rather than at the maximum emission of pure rotational $\mathrm{H_2}$ levels.
   We detect five levels of $\mathrm{HD},\, v=1$ from which we can derive an excitation temperature around $T_{\rm ex} \sim 480 - 710\, \mathrm{K}$. Due to high continuum in the Orion Bar, fringes lead to high noise levels beyond $15\,\mu\mathrm{m}$, no pure rotational lines of $\mathrm{HD}$ are detected. The comparison to PDR models shows that a range of thermal pressure $P = (3-9)\times 10^7\,\mathrm{K}\,\mathrm{cm}^{-3}$ with no strong constraints on the intensity of the UV field $G_0$ are compatible with $\mathrm{HD}$ observations. This range of pressure is compatible with previous estimates from $\mathrm{H}_2$ observations with JWST.}
   {This study provides a new detailed analysis of
   $\mathrm{HD}$ formation and excitation in PDRs. State-of-the-art PDR models with parameters best reproducing other tracers' emission are compatible with $\mathrm{HD}$ observations, highlighting 
   the coherence of the different studies. This is also the first time that observations of $\mathrm{HD}$ emission lines in the near-infrared are used to put constraints on the thermal pressure in the PDR, even though the lines are very faint.}


   \maketitle


\section{Introduction}

Hydrogen deuteride ($\mathrm{HD}$) was first detected through ultraviolet absorption in its first two rotational levels towards the diffuse interstellar cloud located in front of the bright star $\zeta$ Ophiuchii with the Copernicus satellite \citep{1979ApJ...227..483W}. Further ultraviolet observations towards less bright stars have been conducted thanks to FUSE (Far Ultraviolet Space Explorer), allowing additional $\mathrm{HD}$ detections \citep{2000ApJ...538L..69F}.
The $\mathrm{HD}$ infrared emission through its lowest pure rotational transition of $J = 1\rightarrow0$ at $112\,\mu\mathrm{m}$ has been later obtained with the Long Wavelength Spectrometer (LWS) on board the Infrared Space Observatory (ISO) and with the Photodetector Array Camera and Spectrometer (PACS) on board Herschel, towards the Orion Bar \citep{Wright:1999,Joblin_2018}, a prototypical highly irradiated photodissociation region (PDR) 
\citep[for a review see][]{Hollenbach_1997,Wolfire_2022}, located in the Orion Nebula and by \citet{Bertoldi:1999} in the Orion Molecular outflow. That same transition was also observed in absorption towards W49 by \citet{Caux_2002}, revealing the presence of a cold molecular cloud with extreme depletion on the line of sight.
A detection of its ro-vibrational emission (1-0 R(5)) in the Orion region has been first reported by \citet{2002ApJ...566..905H} with the United Kingdom Infrared Telescope (UKIRT), and attributed to shock excitation. However, a single line was observed, which precludes a detailed analysis.
The InfraRed Spectrograph (IRS) on board the Spitzer Space telescope allowed to detect rotationally excited $\mathrm{HD}$ emission in different shocked Herbig-Haro objects \citep{2012ApJ...753..126Y}.
Very recently, the James Webb Space Telescope (JWST), with its high spatial resolution and high sensitivity, enabled the detection of several emission rotational lines of $\mathrm{HD}$, originating from the ground vibrational level in outflow regions across the galactic disk \citep{Francis:2025}.

\begin{figure}
    \centering
    \includegraphics[width=\linewidth]{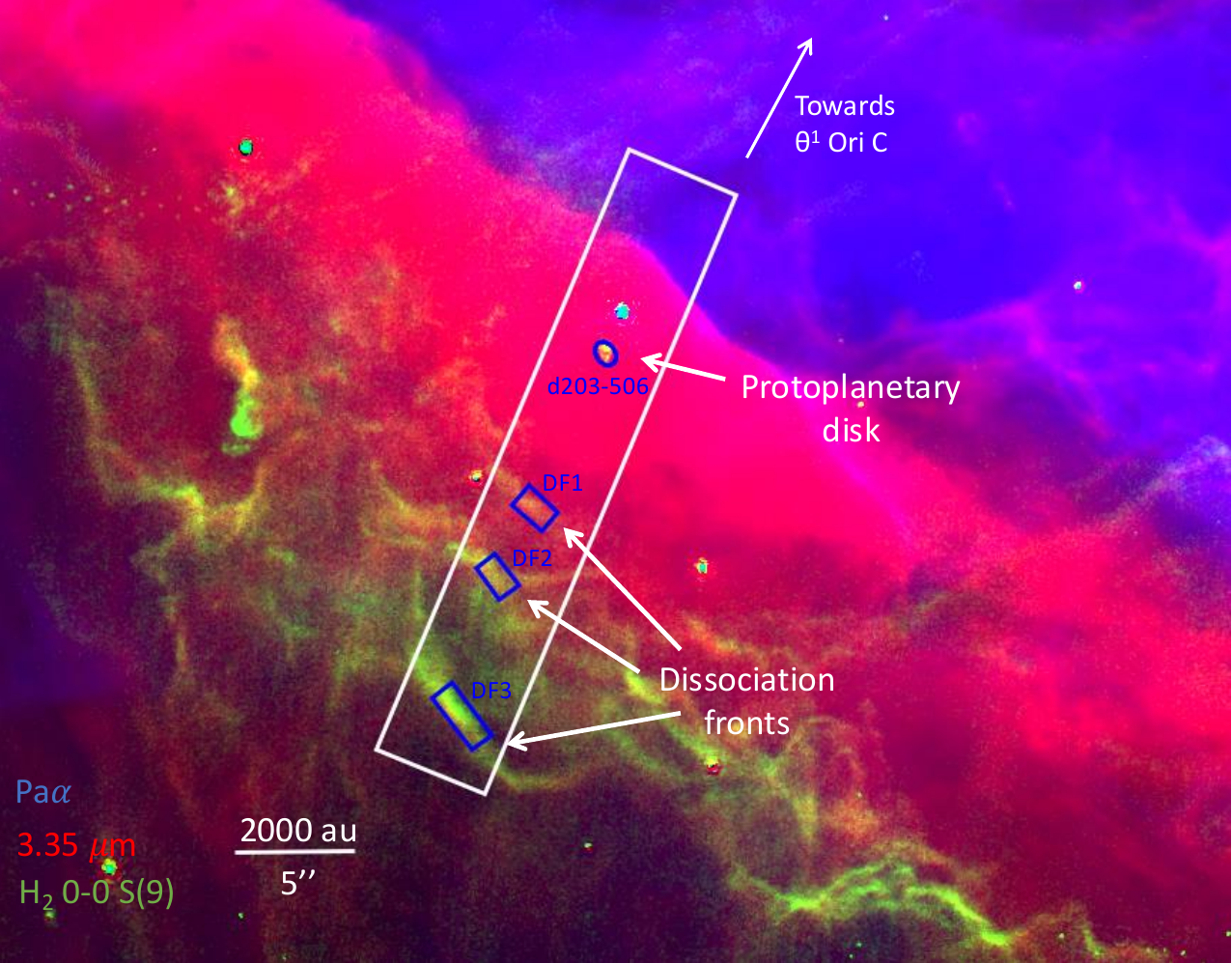}
    \caption{JWST NIRCam composite image of the Orion Bar, located in the Orion molecular cloud \citep{2024A&A...685A..73H}. Red is the $3.35 \,\mu\mathrm{m}$ emission (F335M filter), blue is the emission of Pa$\alpha$ (F187N filter subtracted by F182M filter) and green is the emission of the $\mathrm{H}_2$ 0–0 S(9) line at $4.70\,\mu\mathrm{m}$ (F470N filter subtracted by F480M filter). The white box represents the Field of view of NIRSpec. The blue boxes correspond to the aperture where the spectra are averaged in the three dissociation fronts (DF1, DF2 and DF3).}
    \label{fig:rgb_OB}
\end{figure}

Several rovibrational $\mathrm{HD}$ transitions have also been reported by \cite{2024A&A...685A..74P} towards the Orion Bar. Fig. \ref{fig:rgb_OB} displays an RGB view of this region made with near infrared camera (NIRCam) images. This region, located at $d = 414$ pc \citep{Menten_2007}, acts as an interstellar laboratory to study the emission of $\mathrm{HD}$ because the conditions of irradiation enable  high excitation of this molecule and thus its detection. Indeed, the Orion Bar is exposed to the intense FUV field from the Trapezium cluster, which is dominated by the O7-type star $\theta^1$ Ori C which has an effective temperature of $T_{\rm eff} \simeq 40{,}000\,\mathrm{K}$. The intense FUV radiation field incident on the ionization front (IF) of the Bar is estimated to be $G_0 = 2-7 \times 10^4$ as derived from FUV-pumped IR-fluorescent lines of $\mathrm{O\,I}$ and $\mathrm{N\,I}$ \citep{2024A&A...685A..74P}. The detection of several $\mathrm{HD}$ rovibrational lines in the Orion Bar offers the opportunity to study deuterium chemistry in this high radiation field environment and cross-check the physical conditions derived within the Bar from its excitation compared to other molecular probes.

Early detailed computations of $\mathrm{HD}$ excitation into a model studying the atomic to molecular transition of $\mathrm{D}/\mathrm{HD}$ due to photodissociation were done by \citet{1986ApJS...62..109V,1988A&A...190..215V}, in order to interpret the UV interstellar $\mathrm{HD}$ absorption spectra obtained by Copernicus. \citet{2002A&A...390..369L} extended the physical model to interpret the FUSE observations \citep{Lacour_2005}.
However, only rotational excitation within $v = 0$ had been detected and was taken into account in the models. 

In the present paper, we present a comprehensive treatment of $\mathrm{HD}$ chemistry and excitation in the frame of the Meudon PDR code \citep{Lepetit:2006,Goicoechea_2007,2008A&A...485..127G,Le_Bourlot_2012,Bron_2014} using newly available molecular data involving vibrationally excited $\mathrm{HD}$. These are described in Sect. \ref{Sect:HD_excitation}. The spatial variation of the physical parameters and the associated $\mathrm{HD}$ formation/excitation mechanisms are outlined for an isobaric chemical PDR model in Sect. \ref{Sec:model_Results}. Observational data from the PDRs4All program are presented in Sect. \ref{Sec:Observations}. Detection of $\mathrm{HD}$ and analysis of its excitation in the dissociation fronts is also presented in this section. We compare the observations to our model results in  Sect. \ref{Sec:Fit_Obs} and report our conclusions in Sect. \ref{Sec:Conclusions}.

We emphasize that we do not discuss the elemental $\mathrm{D/H}$ ratio that would require the full analysis of $\mathrm{H}_2$. Nevertheless, we demonstrate how a detailed analysis of the excitation allows to shed light on the physical processes at work.

\section{$\mathrm{HD}$ chemistry and excitation}\label{Sect:HD_excitation}

The Meudon PDR code solves 
the chemical and thermal balance of a 
plane parallel slab of gas submitted to an external radiation field. A detailed radiative transfer treatment allows to compute the photodissociation probabilities of several molecules, including $\mathrm{H_2}, \mathrm{HD}, \mathrm{CO}$ from the integration of the state dependent photodissociation cross-sections multiplied by the intensity of the radiation field computed at each point of the 1D structure, accounting for dust and gas contributions, as explained in the previously cited papers. Here, we infer the statistical equilibrium of the first $52$ levels of $\mathrm{HD}$ that encompass the $v = 2,\, J = 11$ level (highest energy levels at $E = 17402.7\,\mathrm{K}$) within the $X$ electronic ground state. The equations are very similar to those describing  $\mathrm{H}_2$ excitation, including $v,J$ dependent 
photodissociation rates resulting from discrete line absorption followed by emission into the continuum of the $X$ electronic ground state as computed by \cite{2006A&A...445..361A}. The general form of the detailed balance equations can be written as
\begin{equation}
n_{i}\,\left(\sum_{j\neq i} R_{ij}+D_{i}\right)=\sum_{j\neq i} n_{j}\,R_{ji}+F_{i},
 \label{Eq:Steady_state}
\end{equation}
where $n_i$ is the population of the molecule on level $i$ (in $\mathrm{cm^{-3}}$), $D_i$ (in $\mathrm{s}^{-1}$) are chemical destruction terms (including photodissociation), $R_{ij}$ and $R_{ji}$ (in $\mathrm{s}^{-1}$) are the transition rates from level $i$ to level $j$ (respectively from level $j$ to level $i$) including radiative and collisional processes and $F_i$ (in $\mathrm{cm}^{-3}\, \mathrm{s}^{-1}$) is the formation rate of $\mathrm{HD}$ into level $i$. We detail below the main new features of our model.

\subsection{Formation and destruction processes}\label{Sec:HD_Chemistry}

\subsubsection{Main reactions}
 
We consider both gas phase and grain surface reactions that contribute to $\mathrm{HD}$ formation, including Langmuir Hinshelwood and Eley-Rideal mechanisms, similarly as $\mathrm{H_2}$ \citep{Le_Bourlot_2012}. We follow \citet{2011ApJ...737...44G} to build the gas phase chemistry of $\mathrm{HD}$ except for the photodissociation mechanism that we compute from the individual photodissociative absorption radiative transitions previously computed by \citet{2006A&A...445..361A}. $\mathrm{HD}$ chemical formation is mainly achieved through reactions of different deuterium containing species with $\mathrm{H_2}$ and destruction  is dominated by gas phase reactions with various forms of hydrogen, in addition to photodissociation.

\begin{itemize}
    \item Formation:
    \begin{align}
       \mathrm{D} + \mathrm{H}_2 &\rightarrow  \mathrm{HD}  + \mathrm{H} & E_a= -3820\, \mathrm{K}\label{Eq:HD_F1}\\
       \mathrm{D}^+ + \mathrm{H}_2   &\rightarrow \mathrm{HD} + \mathrm{H}^+ \label{Eq:HD_F2}\\
       \mathrm{H_2D}^+  + \mathrm{H}_2   &\rightarrow \mathrm{HD} + \mathrm{H}_3^+ & \Delta H =-232\, \mathrm{K} \label{Eq:HD_F3}\\
       \mathrm{D} + \mathrm{H\!::} &\rightarrow \mathrm{HD} \label{Eq:HD_F4}
    \end{align}
    \item Destruction:
    \begin{align}
       \mathrm{HD} + \mathrm{H} &\rightarrow \mathrm{H}_2 + \mathrm{D} & E_a =4240\, \mathrm{K}\label{Eq:HD_D1}\\
       \mathrm{HD} + \mathrm{H}^+ &\rightarrow \mathrm{H}_2 + \mathrm{D}^+ & \Delta H =-464\, \mathrm{K}\label{Eq:HD_D2}\\
       \mathrm{HD} + \mathrm{H}_3^+ &\rightarrow \mathrm{H}_2 + \mathrm{H_2D}^+ \label{Eq:HD_D3} \\
       \mathrm{HD} + h\nu &\rightarrow \mathrm{H} + \mathrm{D} \label{Eq:HD_D4}
    \end{align}
\end{itemize}
The negative difference of enthalpy variations corresponds to endothermic pathways. Reactions \ref{Eq:HD_F1} and \ref{Eq:HD_D1} possess activation barriers $E_a$ that largely overcome the involved enthalpy variations ($|\Delta H| =420 K$).
$\mathrm{H\!::}$ in reaction \ref{Eq:HD_F4} stands for $\mathrm{H}$ chemisorbed on grains \citep{Le_Bourlot_2012}, that may react with atomic gas phase deuterium impinging on the chemisorbed site.
The neutral-neutral reactions \ref{Eq:HD_F1} and \ref{Eq:HD_D1}  are specifically taking place in the warm irradiated front of dense and strong PDRs where the high temperature and/or the high excitation of $\mathrm{H_2}$ allows to overcome the energy barriers. The ion-molecule reactions occur principally in the more shielded, dense and colder regions and are driven by cosmic rays. Photodissociation is the main destruction route of $\mathrm{HD}$ over an extended range of the cloud.

The couples of reactions (\ref{Eq:HD_F1}-\ref{Eq:HD_D1}, \ref{Eq:HD_F2}-\ref{Eq:HD_D2}, \ref{Eq:HD_F3}-\ref{Eq:HD_D3}) involve forward and  reverse competing paths that depend on the excitation state of $\mathrm{H_2}$. A more complete discussion is given in Sect. \ref{Sub:HD_Abundance} and \ref{Sub:neutral}.

\subsubsection{State-to-state chemical reactions}

The detailed balance equations governing the excitation of a molecule also include chemical formation and destruction processes, that depend on the internal states of the reactants and products. This dependence is usually unknown and is either neglected or treated with educated guesses, which may lead to significant errors in the resulting computed populations and lines intensities.

The two neutral-neutral reactions \ref{Eq:HD_F1} and \ref{Eq:HD_D1} have been studied for different rovibrational states of $\mathrm{H_2}$ and $\mathrm{HD}$ by \cite{2019JChPh.150h4301B} and \cite{10.1063/5.0017697} in a state-to-state quasiclassical approach and we have introduced these data in our model\footnote{Both $\mathrm{H_2}$ and $\mathrm{HD}$ excitation balance is affected by these reactions.}.
The rate coefficient for reaction~\ref{Eq:HD_F1} includes 261 levels of $\mathrm{H_2}$ and $398$ levels of $\mathrm{HD}$. The rate coefficients for the reverse reaction~\ref{Eq:HD_D1}, that involves an activation barrier, uses $301$ levels of $\mathrm{H_2}$ and $239$ levels of $\mathrm{HD}$.
They provide a remarkable update to the 
analytic global rate previously recommended  by \cite{2011ApJ...737...44G}.

We have introduced these data in the PDR code for both $\mathrm{H}_2$ and $\mathrm{HD}$ detailed balance equations and chemical steady state computation. The impact of using these state-to-state rate coefficients is examined in Appendix \ref{Sub:neutral}. We also account for the dependence on the ortho-to-para ratio of $\mathrm{H}_2$ for reaction \ref{Eq:HD_F3}  that becomes significant in the shielded cold part of the cloud. For all other $\mathrm{HD}$ formation reactions, we suppose that the newly formed molecules are distributed proportionally to a Boltzmann factor at the gas temperature.

\subsection{Excitation/de-excitation processes}

\subsubsection{Radiative rovibrational transitions within $X$}

A significant difference between $\mathrm{HD}$ and $\mathrm{H_2}$ is that weak electric dipolar transitions within $X\,^1\Sigma^+_g$ 
may occur, due to a small dipole moment  ($8.4\times 10^{-4}\,\mathrm{Debye}$), resulting from the
difference of location of the center of mass and the center of charge of the molecule.
Energy levels and radiative transition probabilities of $X$ are taken from \citet{2019ApJ...878...95A}, as found on the HITRAN\footnote{\href{https://hitran.org}{https://hitran.org}} database, updating previous calculations by \cite{Abgrall:1982}.

\subsubsection{Ultraviolet electronic radiative transitions}\label{Sec:Radiative}

We include the transitions linking the ground state to the
four upper electronic states $B\,^1\Sigma_u$, $C\,^1\Pi_u$, $B'\,^1\Sigma_u$ and $D\,^1\Pi_u$, where discrete line absorption is followed by discrete and continuous emission. The discrete emitted transitions towards rovibrationally excited levels within the ground state are followed by subsequent infrared radiative cascades that may be detected. On the other hand, transitions toward the continuum lead to photodissociation of the molecule. The formalism closely follows that of $\mathrm{H}_2$, as emphasized previously.
\subsubsection{Collisions}\label{Sec:Collisions}

Different colliders ($\mathrm{H}$, $\mathrm{He}$, $\mathrm{H}_2$, $\mathrm{H}^+$, electrons) may induce collisional excitation/de-excitation between rovibrational levels belonging to the ground electronic state of $\mathrm{HD}$. We have collected the different available collisional data.

Rovibrational collisional excitation rate coefficients with $\mathrm{H}$ are taken from the recent computations of \citet{2022MNRAS.513..900D}. They are provided up to level $v = 2,\, J = 11$ of $\mathrm{HD}$. The rates for collisions with $\mathrm{He}$ are taken from \citet{2012ApJ...744...62N}, as found in the UGAMOP \footnote{\href{https://sites.physast.uga.edu/ugamop/}{https://sites.physast.uga.edu/ugamop/}} database. They include $872$ transitions from levels up to $v = 17,\, J = 1$, but with low $J$ only. 
The corresponding data for collisions with $\mathrm{H}_2$ are taken mostly from \citet{1999MNRAS.309..833F} except for levels with $v = 0,\,J < 9$ where recent results from \citet{2019MNRAS.488..381W} are used. Collision rates with $\mathrm{H}^+$ come from \citet{2022JChPh.157u4302G}, as found in the EMAA\footnote{\href{https://emaa.osug.fr/details/HD}{https://emaa.osug.fr/details/HD}} database. Note that many pairs of levels are still missing from these data.

We also consider collisions with electrons that have been approximated 
by \citet{1975JPhB....8.2846D,1977A&A....54..645D}. The suggested expression of the rate coefficient $\alpha$ in $\mathrm{cm}^3\,\mathrm{s}^{-1}$ for low dipole moment values is reported as\footnote{Used within $v=0$ only.}
\begin{align}
     \alpha\left(j\rightarrow j+1\right) &= \frac{3.56\times 10^{-6}\,D^{2}}{\sqrt{T}\,\left(\frac{2\,j+1}{j+1}\right)} \, \exp\left(-\beta\,\Delta E\right) \nonumber \\
     &\times \ln\left(C\,\Delta E+\frac{C}{\beta}\,\exp\left(\frac{-0.577}{1+2\,\beta\Delta E}\right)\right)
     \label{Eq:HD_e-}
\end{align}
with:
\[
C=\frac{9.08\times 10^{3}}{B_{0}\,\left(j+1\right)}  ;\quad \beta=\frac{11600}{T} ;\quad\Delta E=2.48\times 10^{-4}\,B_{0}\,\left(j+1\right) 
\]
\[
D=8.4\times 10^{-4}\,\mathrm{Debyes}\,;\quad B_{0}=45.655\,\mathrm{cm}^{-1}
\]
$\Delta\mathrm{E}$ is an energy term expressed in $\mathrm{eV}$ whereas $\mathrm{C}$ and $\mathrm{\beta}$ are homogeneous to the inverse of energy, expressed in $\mathrm{eV}^{-1}$. Note that Eq.\,\ref{Eq:HD_e-} leads to huge variations of $\alpha$ for $T < 100\,\mathrm{K}$, that are not critical due to their negligible contribution.

Given the different availabilities, we finally solve the detailed balance equations \ref{Eq:Steady_state} for the first $52$ levels of the $X$ ground state, where the highest energy level is $v = 2,\, J = 11$, at $17402.7\, \mathrm{K}$ above the ground level. 
The levels located above that energy are assumed to depend only on radiative processes that are readily computed, following the cascade formalism detailed in \citet{1986ApJS...62..109V,1988A&A...190..215V}.

\section{A fiducial model}\label{Sec:model_Results}

\subsection{Parameters}\label{Parameters}

We use version 7 of the Meudon PDR code\footnote{Available at
\href{https://pdr.obspm.fr}{https://pdr.obspm.fr}}, modified to include all processes described in Sect. \ref{Sect:HD_excitation}. Ultraviolet Lyman series absorption lines from $\mathrm{H}$, $\mathrm{D}$, as well as Lyman (B-X), Werner (C-X), (B'-X) and (D-X) transitions of $\mathrm{H}_2$ and $\mathrm{HD}$ are included in the radiative transfer model, as described in \citet{Goicoechea_2007}. Hence all incidental line overlap between these species and self-shielding effects are accounted for straight away. We adopt a base model of the Orion Bar environment following previous PDRs4All publications as displayed in Table\,\ref{tab:Reference-model-parameters.}. Note that a constant pressure model is compatible with steady state.

The cloud is illuminated by the standard isotropic Interstellar Radiation Field (ISRF) plus the O7 type $\theta$ Orionis irradiating star \citep{2023A&A...673A.149H}, set at a distance $d_* = 0.5 \,\mathrm{pc}$ from the edge of the cloud \citep[based on values given in][]{Pellegrini_2009,Odell_2020}. We model the star spectrum with the Castelli \& Kurucz atlas\footnote{\href{https://www.stsci.edu/hst/instrumentation/reference-data-for-calibration-and-tools/astronomical-catalogs/castelli-and-kurucz-atlas}{https://www.stsci.edu/hst/instrumentation/reference-data-for-calibration-and-tools/astronomical-catalogs/castelli-and-kurucz-atlas}} \citep{2003IAUS..210P.A20C} for a star with an effective temperature of $T_{\rm eff} = 4 \times 10^4\,\mathrm{K}$ and $\log g = 4.5$. The effective $G_0$ at the cloud edge, with respect to the standard Habing ISRF reference \citep{Habing:1968}, is 1.4 10$^4$ $G_0$, in the range of previously suggested values \citep{Joblin_2018}. Table \ref{tab:d-G0.} and 
Fig. \ref{Fig:OBHD_G0_d} give the correspondence between $G_0$ and $d_*$ for different values of $d_*$.
The extinction curve is taken from \citet{1988ApJ...328..734F}. 
We take the power law suggested by \citet{1977ApJ...217..425M} to describe the dust size distribution. The grains composition is a mixture of carbon and silicates as described by \citet{2001ApJ...548..296W}.  
The adopted Line of Sight (LoS) is HD 38087 \citep{Joblin_2018}, with $R_V = 5.5$, as prescribed in \citet{2024A&A...685A..73H}.
We use a relative abundance $\delta_D = \mathrm{D}/\mathrm{H} = 1.5\times 10^{-5}$, representative of the standard elemental ratio \citep{2006ApJ...647.1106L}. A minimal oxygen surface chemistry is included, following \citet{2009ApJ...690.1497H}.

\begin{table}
\caption{
Fiducial model parameters.\protect\label{tab:Reference-model-parameters.}}
\centering%
\begin{tabular}{cc}
\hline\hline
Parameters & Value \\
\hline
 $P$ $(\mathrm{K}\,\mathrm{cm}^{-3})$ & $5\times 10^7$\\
 $d_*$ $(\mathrm{pc})$& $0.5$ \\ 
 $G_0$& $1.37 \times 10^4$\\
 $A_{\mathrm{V}}$ & $30$\\
 $\zeta$ $(\mathrm{s}^{-1})$& $10^{-16}$ \\
 $v_{t}$ $(\mathrm{km}\,\mathrm{s}^{-1})$  & $2$\\
 $R_\mathrm{V}$ &  $5.5$  \\
 $\frac{N_\mathrm{H}}{E_{\mathrm{B-V}}}$ $(\mathrm{cm}^{-2})$  & $1.57\times 10^{22}$ \\
 Dust to gas mass ratio &  $10^{-2}$ \\ 
 Grain distribution  & proportional to $a ^{-3.5}$ \\
 minimum grain radius $a_{min}$ & $0.003\, \mu\mathrm{m}$ \\
 maximum grain radius $a_{max}$ &  $0.3\, \mu\mathrm{m}$ \\
\hline
\end{tabular}
\tablefoot{$A_V$ stands for the total visual extinction of 
the PDR. $\zeta$ stands for the cosmic ionization rate of $\mathrm{H_2}$}
\end{table}

\subsection{Structure}\label{sect:structure}

\begin{figure}
\centering\includegraphics[width=1.05\columnwidth]{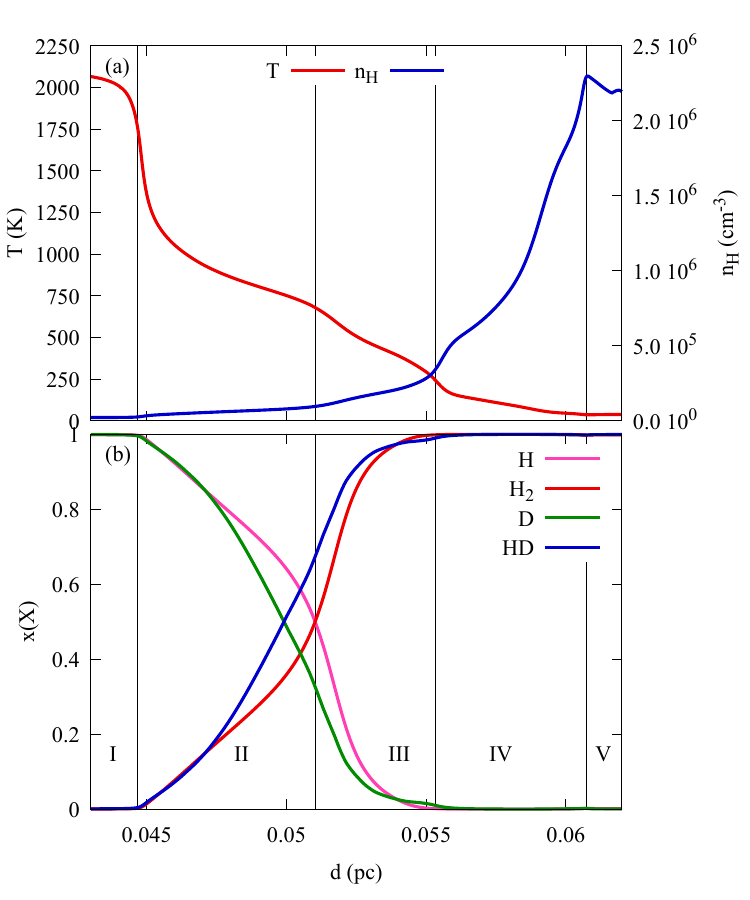}
\caption{(a) Temperature (red line, left axis) and density profiles (blue line, right axis) as a function of distance from the ionization front, $d$, for the fiducial model. 
See Sect. \ref{sect:structure} for the definitions of zones $I$ to $V$. 
(b) Relative abundances of $\mathrm{H}$ and $\mathrm{H}_2$, (resp. $\mathrm{D}$ and $\mathrm{HD}$), normalized by $n(\mathrm{H}) + n(\mathrm{H}_2)$ (resp. $n(\mathrm{D}) + n(\mathrm{HD})$). Values at the transitions are given in Table\,\ref{tab:d-T-n-Values.}.}
\label{Fig:OBHD_zones_P5e7_d5}
\end{figure}

Fig. \ref{Fig:OBHD_zones_P5e7_d5} shows the temperature and density profiles of the relevant regions of the cloud edge. Five zones, corresponding to five different regimes are defined and used subsequently, as described below.

\begin{enumerate}
    \item Zone $I$ extends from the edge of the cloud to an abrupt drop in the temperature profile. As shown below (Sect. \ref{Heating}), this is the point where $\mathrm{H_2}$ collisional excitation followed by radiative decay turns from a heating to a cooling process when collisions become more efficient due to the higher density.
    
    \item Zone $II$ is partially atomic, and exhibits a rising amount of molecular gas up to the location where atomic and molecular hydrogen abundances are equal.

    Efficient self-shielding in $\mathrm{H}_2$ absorption transitions is taking place and allows some molecules to form.
    This zone extends also to the point where atomic and molecular deuterium abundances become equal.

    \item Zone $III$ starts at the position where $n(\mathrm{H_2}) = n(\mathrm{H})$, beyond which the gas is mostly molecular. 

    \item Zone $IV$ begins at the transition from $\mathrm{C}^+$ to $\mathrm{CO}$ as the dominant carbon bearing species. It marks the beginning of the molecular region of the PDR for all species.
    Deuterium exchange reactions   $ \mathrm{HD} + \mathrm{H}_3^+  \rightleftarrows \mathrm{H_2D}^+ + \mathrm{H}_2$ take place 
    in zone $IV$, allowing  formation of further deuterated compounds ($\mathrm{DCO^+}$, ...).
    
    \item Zone $V$ is a high density, low temperature, dark molecular environment where reactions in ice mantles play a dominant role. It is not relevant for our study.
\end{enumerate}
 
The sizes of zones $I$, $II$, $III$ and $IV$ are
respectively $d_I = 4.5\, 10^{-2}$, $0.64 \, 10^{-2}$, $0.43 \, 10^{-2}$ and $0.54\, 10^{-2}\,\mathrm{pc}$, (equivalently  $\sim \, 9200$, $1300$, $900$ and $1100 \ \mathrm{au}$),
with the parameters reported in Table \ref{tab:Reference-model-parameters.}.

\begin{table}
\caption{Transition values.}
\protect\label{tab:d-T-n-Values.}
\centering%
\begin{tabular}{ccccc}
\hline\hline
 & $I\,/\,II$ & $II\,/\,III$ & $III\,/\,IV$ & $IV\,/\,V$\tabularnewline
\hline
$d\,\left(\mathrm{pc}\right)$ & $0.045$ & $0.051$ & $0.055$ & $0.061$\tabularnewline
$T\,\left(\mathrm{K}\right)$ & $1750$ & $679$ & $240$ & $36$\tabularnewline
$n_{\mathrm{H}}\,\left(\mathrm{cm}^{-3}\right)$ & $2.6\,10^{4}$ & $9.6\,10^{4}$ & $3.5\,10^{5}$ & $2.3\,10^{6}$\tabularnewline
\hline
\end{tabular}
\end{table}

\subsection{Heating and cooling processes}\label{Heating}

\begin{figure}
\centering\includegraphics[width=1\columnwidth]{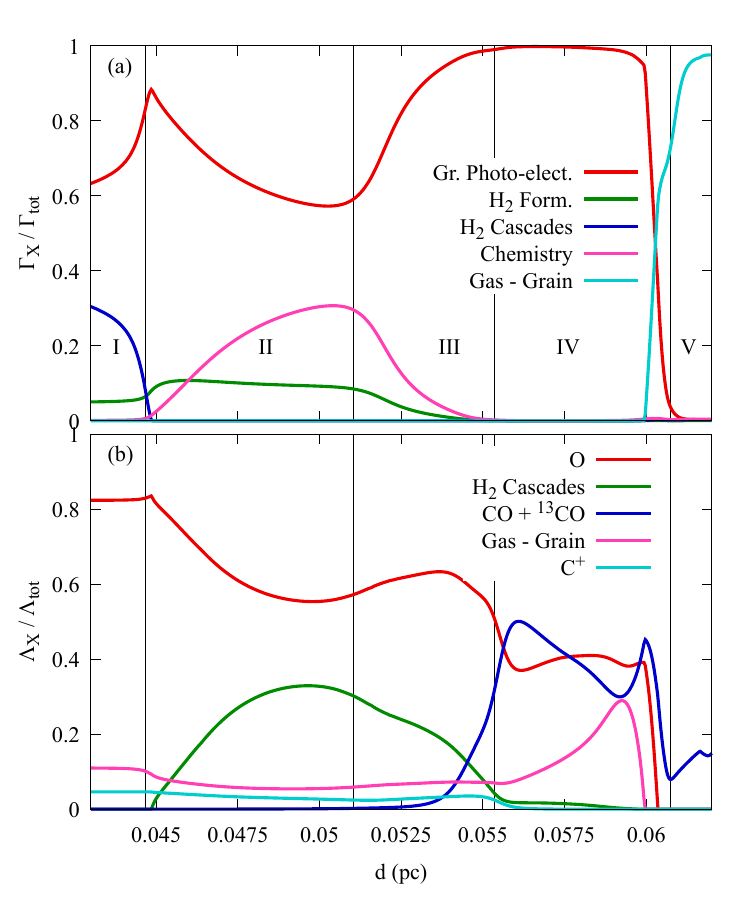}
\caption{Main contributions to (a) heating ($\Gamma$) and (b) cooling ($\Lambda$) of the gas as a function of distance $d$ from the ionization front. Each heating (resp. cooling) term is normalized by the total heating/cooling.}
\label{Fig:OBHD_Heat_Cool_P5e7_d5}
\end{figure}

Fig. \ref{Fig:OBHD_Heat_Cool_P5e7_d5} shows the main heating and cooling processes of the gas. We assume thermal balance, so the total heating and cooling rates are equal. Unsurprisingly the main heating process at the cloud edge is photo-electric effect on dust. However, two other processes have a significant contribution, namely radiative cascades and chemical reactivity: 
\begin{itemize}
    \item In zone $I$, the radiative pumping of $\mathrm{H}_2$ followed by collisional deexcitation is important. This may come as a surprise, since the abundance of $\mathrm{H}_2$ is negligible here. But since $\mathrm{H}_2$ photodissociation occurs in discrete transitions, with a dissociating probability of about $10\,\%$,
    nine times out of ten, absorption of an ultraviolet photon is followed by a radiative cascade to an excited vibrational state. 
    The subsequent deexcitation results from a competition between collisional and radiative decay. Collisional de-excitation increases the kinetic energy of the gas and takes place in this front, medium density, warm zone.
    
    \item In zone $II$, self-shielding of $\mathrm{H_2}$ proceeds and collisional excitation followed by radiative emission is now a cooling process (since kinetic energy is transferred into radiation).
    $\mathrm{H}_2$ abundance is also high enough for an active chemistry to take place and  
    exothermic reactions contribute to heating. Appendix\,\ref{Sec:Chem_Heat} details these processes.
\end{itemize}

$\mathrm{H}_2$ formation on grains through the Eley-Rideal process has a minor contribution to gas heating ($\sim 10\%$).
Cooling is dominated by $\mathrm{O}$ far-infrared lines, but here also it is not the only process to participate in the cooling. Everywhere up to the molecular core, gas-grain collisions contribute to about $10\,\%$. Once $\mathrm{H}_2$ survives in a significant amount (zone $II$), it contributes to more than $30\,\%$, and deeper into the cloud, $\mathrm{CO}$ and its isotopes contribute to about $40\,\%$. $\mathrm{C}^+$ has a minor contributions up until the molecular core (less than $4\%$ everywhere). Other species ($\mathrm{C}$, $\mathrm{OH}$, $\mathrm{H_2O}$, ...) are completely negligible. 

\begin{figure*}
\centering\includegraphics[width=1\linewidth]{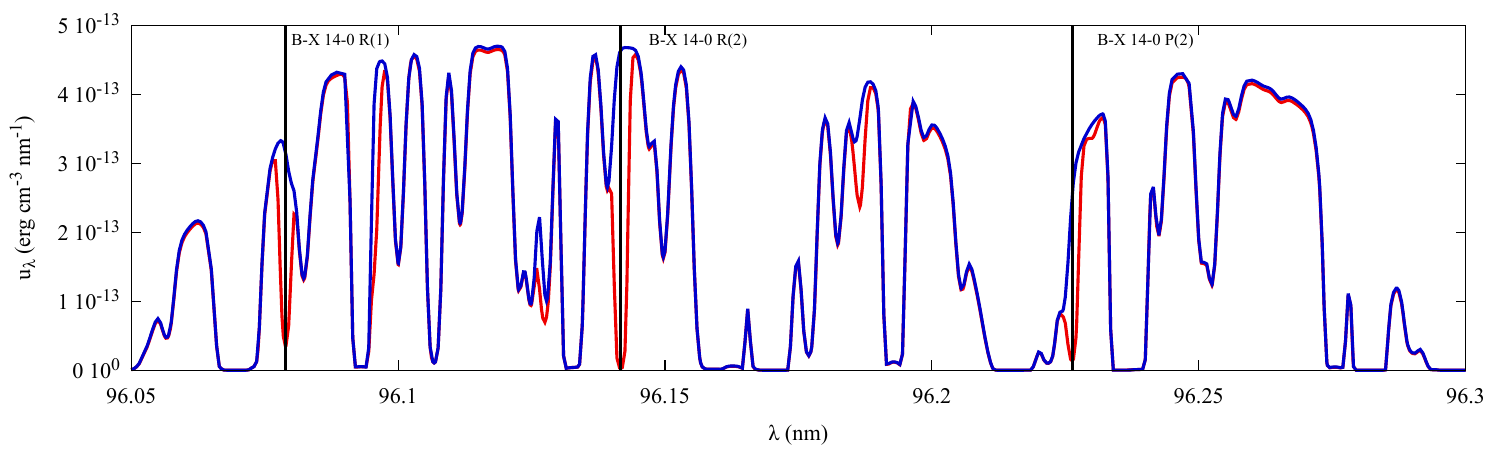}
\caption{Radiative energy density $u_{\lambda}$ at d= 0.05 pc in the $96-96.3\, \mathrm{nm}$ far ultraviolet wavelength range. Blue: $\mathrm{H}_2$ lines, red: $\mathrm{HD}$ lines. The positions of three strong $\mathrm{HD}$ absorption lines are shown using black vertical lines, coming from $v = 0$, $J = 1$ and $J = 2$.}
\label{Fig:OBHD_UV}
\end{figure*}

\subsection{$\mathrm{HD}$ abundance}\label{Sub:HD_Abundance}

\begin{figure}
\centering\includegraphics[width=1\columnwidth]{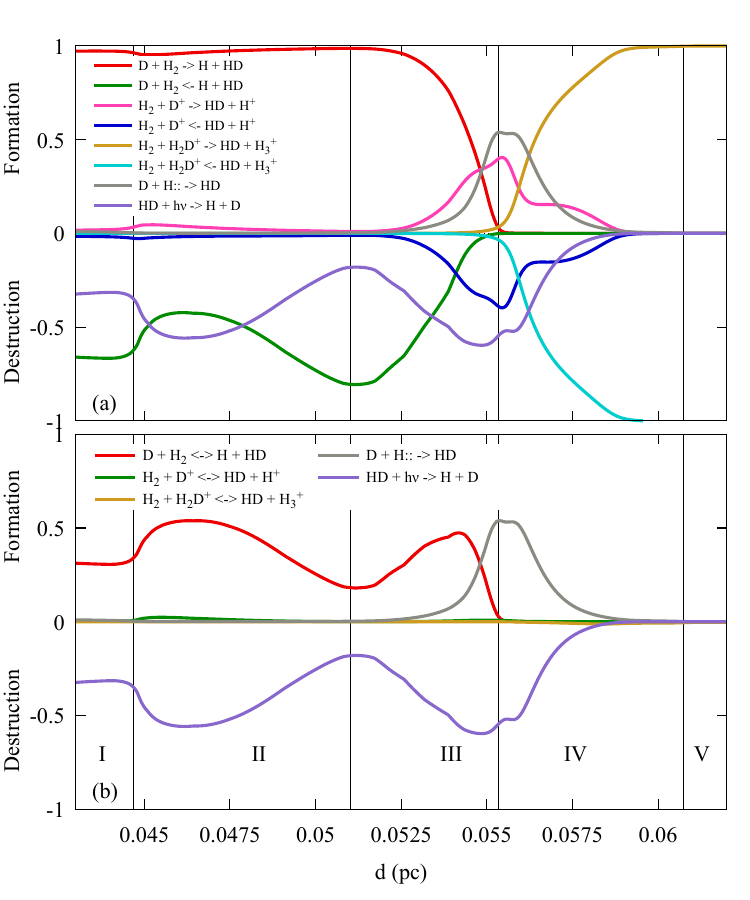}
\caption{Relative contribution of $\mathrm{HD}$ formation and destruction reactions normalized to the total formation rate. (a): individual reactions \ref{Eq:HD_F1} to \ref{Eq:HD_D4}. (b): sum of direct and reverse reactions. The $y$ axis is dimensionless.}
\label{Fig:Rxn_FD_HD_d6}
\end{figure}

$\mathrm{HD}$ formation starts as soon as some molecular $\mathrm{H_2}$ is available, through the $\mathrm{D} + \mathrm{H}_2$ reaction.
Fig. \ref{Fig:Rxn_FD_HD_d6} shows the relative role of reactions \ref{Eq:HD_F1} to \ref{Eq:HD_D4} normalized by the total formation rate.
In panel (a) individual reactions rates are shown, whether in panel (b) the sum of forward and reverse processes is shown, which helps to identify the main mechanisms.

Photodissociation is the major destruction channel of 
$\mathrm{HD}$ up to the middle of zone $III$, as shown in panel (b). We carefully consider the different possible overlap of $\mathrm{HD}$, $\mathrm{H}$, $\mathrm{D}$ and $\mathrm{H_2}$ UV transitions as well as self-shielding effects of $\mathrm{H_2}$  and $\mathrm{HD}$ dissociating transitions.
We show an example in Fig. \ref{Fig:OBHD_UV} where  a small portion of the radiative energy density $u_{\lambda}$ is displayed at $A_{\mathrm{V}} = 1.7$ ($d \simeq 0.05\,\mathrm{pc}$, just before zone $III$).
Absorption lines in blue come from $\mathrm{H}_2$. $\mathrm{HD}$ lines are shown in red. Two lines arising from $v = 0,\,J = 2$ are identified.
The $14-0\, R(2)$ line is only mildly shielded,
whereas the $14-0\, P(2)$ line falls in the wing of a strong $\mathrm{H_2}$ line, impacting the dissociation efficiency. We also note the presence of the strong $14-0\, R(1)$ transition in the wavelength range.
Compared to the simple \citet{1979ApJ...227..466F} (FGK) approximation
of the self-shielding, photodissociation is reduced by 10 to 30\% over zone $II$.

Diverse chemical processes occur in the dense zones $III$ and $IV$, where the temperature gradually decreases from $\sim 600\, \mathrm{K}$ down to $50\, \mathrm{K}$. The transition between zone $III$ and $IV$ corresponds to the position where the role of neutral-neutral reactions involving energy barriers is progressively declining whereas ion-molecule reactions become important. The Eley-Rideal $\mathrm{HD}$ formation process on grains dominates over a small spatial range before reactions with $\mathrm{H}_2\mathrm{D}^+$ take over. We note that photodissociation remains significant in zone $III$ and a large part of zone $IV$.
Appendix\,\ref{Sub:neutral} gives more details.

\subsection{$\mathrm{HD}$ excitation}

\begin{figure}
\centering\includegraphics[width=1\columnwidth]{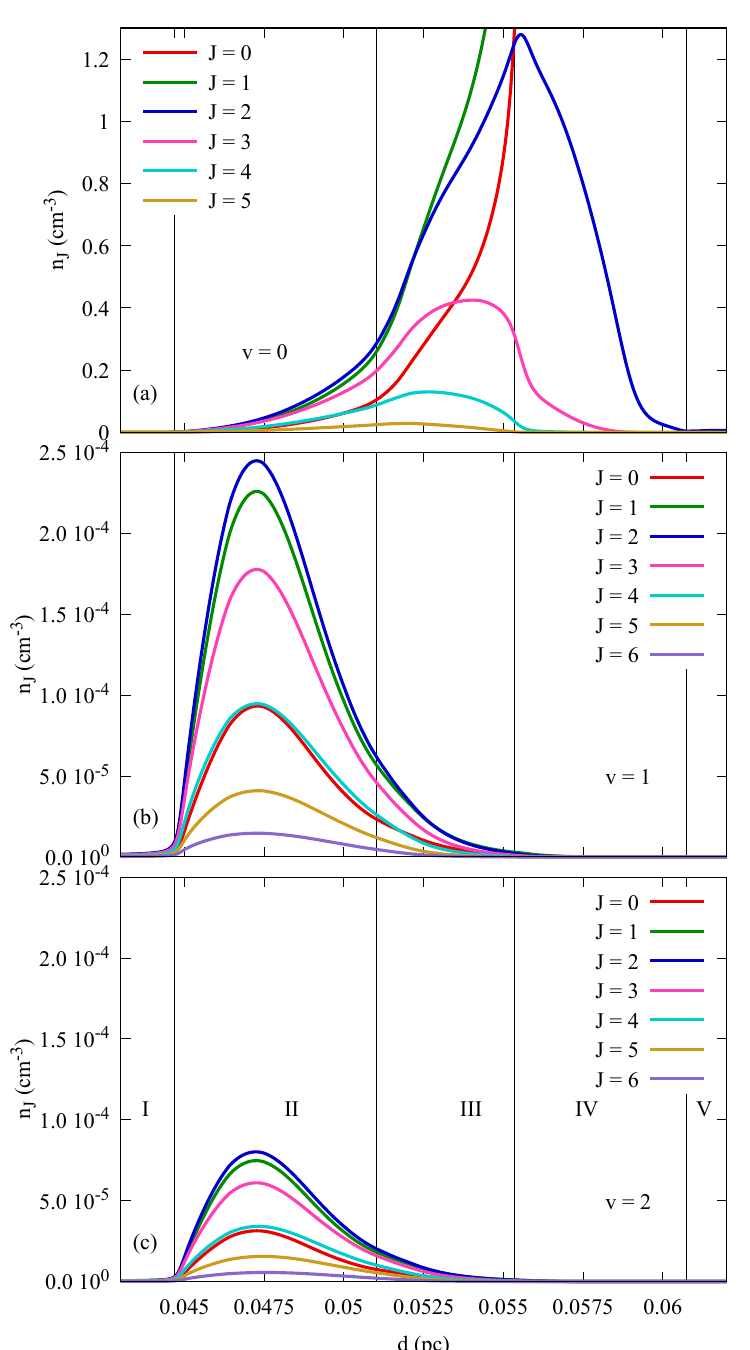}
\caption{
$\mathrm{HD}$ excited rotational levels abundance profiles as a function of distance from the ionization front $d$. (a): $v = 0$ (top panel), (b): $v=1$, (c): $v=2$.
See Fig. \ref{Fig:OBHD_H2v_Lin} for a comparison with $\mathrm{H}_2$.
}
\label{Fig:OBHD_v012_P5e7_d5}
\end{figure}

Fig. \ref{Fig:OBHD_v012_P5e7_d5} shows where, in the PDR, excited levels of $\mathrm{HD}$ are found. Beyond zone $III$ only the lowest levels of the ground vibrational mode $J = 0$ and $1$ are populated. The $v = 0$ levels come mostly from region $III$, while excited vibrational levels come from region $II$, with a well defined separation between both. So $\mathrm{HD}$ vibrational emission comes from a region where $\mathrm{H}_2$ is still a minor component, but is  
subject to enhanced vibrational excitation, as shown in Fig. \ref{Fig:OBHD_H2v_Lin}.

\begin{figure}
\centering\includegraphics[width=1\columnwidth]{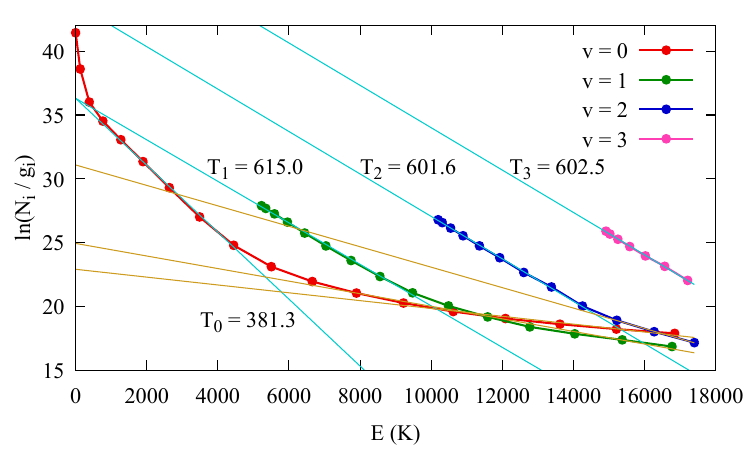}
\caption{Full modeled excitation diagram for $\mathrm{HD}$. Light blue lines: fits to lowest $J$ for each $v$ giving the indicated excitation temperatures $T_{ex}$, light golden lines: fits to high $J$ giving a $T_{ex}^{high\, J}$ beyond the visible change of slope.}
\label{Fig:OBHD_Texcit2_P5e7_d5}
\end{figure}

We compute the full emission spectrum as seen by an external observer, including $\mathrm{HD}$ lines and also all lines from species participating in the cooling of the gas ($\mathrm{H}_2$, $\mathrm{O}$, ...) and included in our model.
$\mathrm{HD}$ $v = 2$ levels are weaker than $v = 1$ levels by about a factor of $3$.
This is confirmed by the excitation diagram shown in Fig. \ref{Fig:OBHD_Texcit2_P5e7_d5}: the $v = 2$ branch is below the $v = 1$ one. As the product $h\nu \times N(J) \times A_{ul}$ (where $A_{ul}$ is the Einstein emission coefficient) is very similar for $v = 2 \rightarrow 1$ transitions compared to $v = 1 \rightarrow 0$ transitions, the difference can be attributed to differences in column densities. See Appendix\,\ref{Sub:Photodissociation} for details and Fig. \ref{Fig:OBHD_Match_HD} for an example spectrum.

An excitation temperature $T_{\rm ex}$ can be derived for each vibrational ladder of the excitation diagram.
To compare these excitation temperatures to the gas kinetic temperature, we compute a "vibrationally weighted" mean gas kinetic temperature $T_{kin}^v$ by:
\begin{equation}
    T_{kin}^v = \int_0^{d_{max}} T(s)\, n_v(s) \, ds / \int_0^{d_{max}} n_v(s) \, ds
\end{equation}
where $T(s)$ is the gas kinetic temperature at position $s$, $n_v(s)$ is the abundance in $\mathrm{cm}^{-3}$ of $\mathrm{HD}$ in vibrational level $v$ (summed over all rotational levels). Results are given in Table\,\ref{tab:Excit_Temper}. The excitation temperatures are close to the "vibrationally weighted" mean gas kinetic temperature for $v=0$, but significantly different for excited $v$.

\begin{table}
\caption{Excitation and vibrationally weighted kinetic temperatures.\label{tab:Excit_Temper}}
\centering%
\begin{tabular}{ccccc}
\hline\hline
 & $v=0,\, J>1$ & $v=1$ & $v=2$ & $v=3$ \\
\hline
 $T_{kin}^v$ & 372 & 897 & 894 & 892 \\
 $T_{\rm ex}$ & 381 & 615 & 602 & 603 \\ 
 $T_{\rm ex}^{\mathrm{high}\,J}$ & 3257 & 2030 & 1249 & - \\
\hline
\end{tabular}
\tablefoot{Vibrationally weighted kinetic temperatures compared to $\mathrm{HD}$ excitation temperatures. All temperatures in $\mathrm{K}$. See Fig. \ref{Fig:OBHD_Texcit2_P5e7_d5} for the fit defining $T_{\rm ex}^{\mathrm{high}\,J}$.}
\end{table}

To understand the difference between $v=0$ and $v>0$ excitation, we explore the different excitation processes by developing the various $R_{ji}$ terms in the detailed balance equations \ref{Eq:Steady_state}.
The relative weight of each process can be assessed by computing the balance of formation and destruction of molecules in a given level $i$ due to each one. We define:
\begin{equation}
    S^e_i = \sum_j x_j\, R^e_{ji}
\end{equation}
where $x_j$ is the relative population of $\mathrm{HD}$ in level $j$. $e$ stands for either $r$: radiative quenching transitions, $c$: collisions transitions, $p$: radiative pumping and cascades  or $f$: formation excitation. The various terms are expressed in $\mathrm{s^{-1}}$.

\begin{figure}
\centering\includegraphics[width=1\columnwidth]{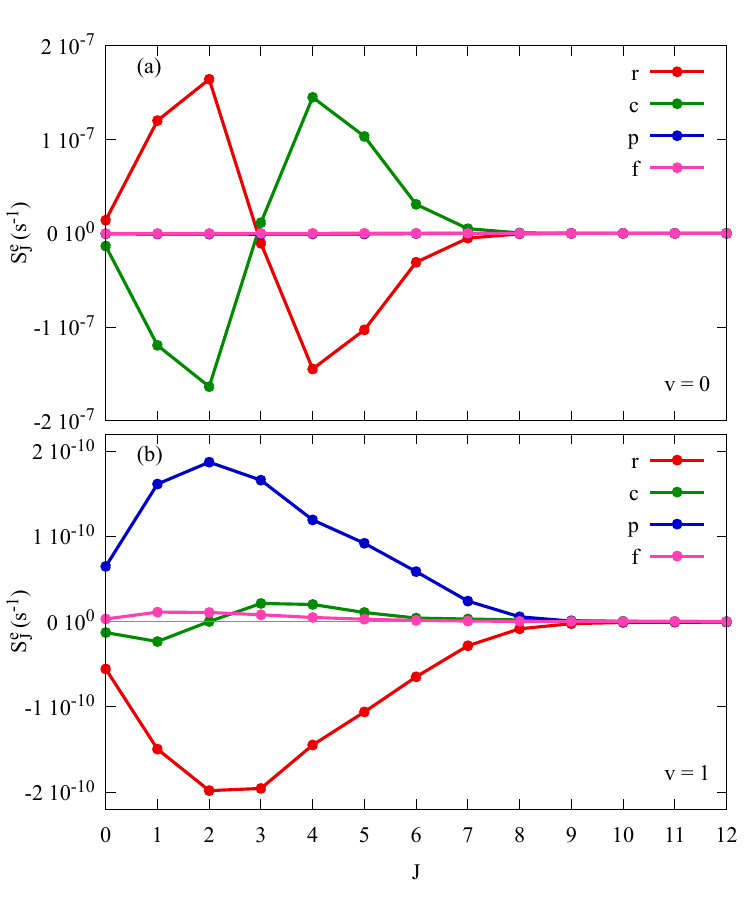}
\caption{Excitation balance $S^e_j$ at $d= 0.05\, \mathrm{pc}$ for (a): $v=0$ and (b): $v=1$. $r$: radiative transitions, $c$: collisions transitions, $p$: pumping and cascades transitions or $f$: formation excitation.}
\label{Fig:OBHD_Excit_HD}
\end{figure}

Fig. \ref{Fig:OBHD_Excit_HD} shows the four processes, at $d = 0.05\, \mathrm{pc}$ for the considered $J$ values belonging to $v=0$ and $v=1$. In the case of $v=0$, the dominant terms in the detailed balance equations \ref{Eq:Steady_state} come from collisions and spontaneous radiative de-excitation. $v=0$ levels peaks in zone $III$, where the density is high enough, so collisions are efficient enough to bring the excitation temperature close to the kinetic temperature. For $v=1$, the populations 
mainly result from radiative pumping balanced by spontaneous radiative
emission. This non-LTE process explains the subthermal excitation ($T_{\rm ex} < T_{kin}^v$)  of vibrational levels of $\mathrm{HD}$. Interestingly, the excitation of $\mathrm{HD}$ due to FUV-pumping is subthermal.
This is due to the small $\mathrm{HD}$ electric dipole moment allowing electric dipolar transitions.
E.g., level $v=1,\,J=10$ of $\mathrm{HD}$ has an inverse radiative lifetime $158$ times higher than the same level of $\mathrm{H}_2$, with the main contribution coming from the electric dipole $R\,(9)$ transition with an Einstein coefficient of $7.65\, 10^{-5}\, \mathrm{s}^{-1}$.

We note also that for each vibrational branch in Fig. \ref{Fig:OBHD_Texcit2_P5e7_d5} the highest $J$ levels follow a higher temperature (last line of Table\,\ref{tab:Excit_Temper}, and brown lines on the figure). This change of slope coincides with the begining of the next $v$ branch, which suggests a coupling between $v$ and $v+1$ levels.
This is confirmed by a careful examination of formation and destruction processes of high $J$ levels: collisional decay from $v+1$ to $v$ with $\Delta J \geqslant 2$ is efficient enough to lead to an overpopulation of these levels compared to populations expected from the cold component.

A list of line intensities is given in Appendix\,\ref{Line_int_fiducial} for rotational transitions, and transitions with $\Delta v = 1$, $2$ and $3$. Most of these lines are not seen in the current spectra (see next Section), but the quoted intensities show which noise level must be reached in order to detect them.

\section{Observations}\label{Sec:Observations}

\subsection{Data reduction}
\label{sec:data_reduc}
\begin{figure*}
    \centering
    \includegraphics[width=\linewidth]{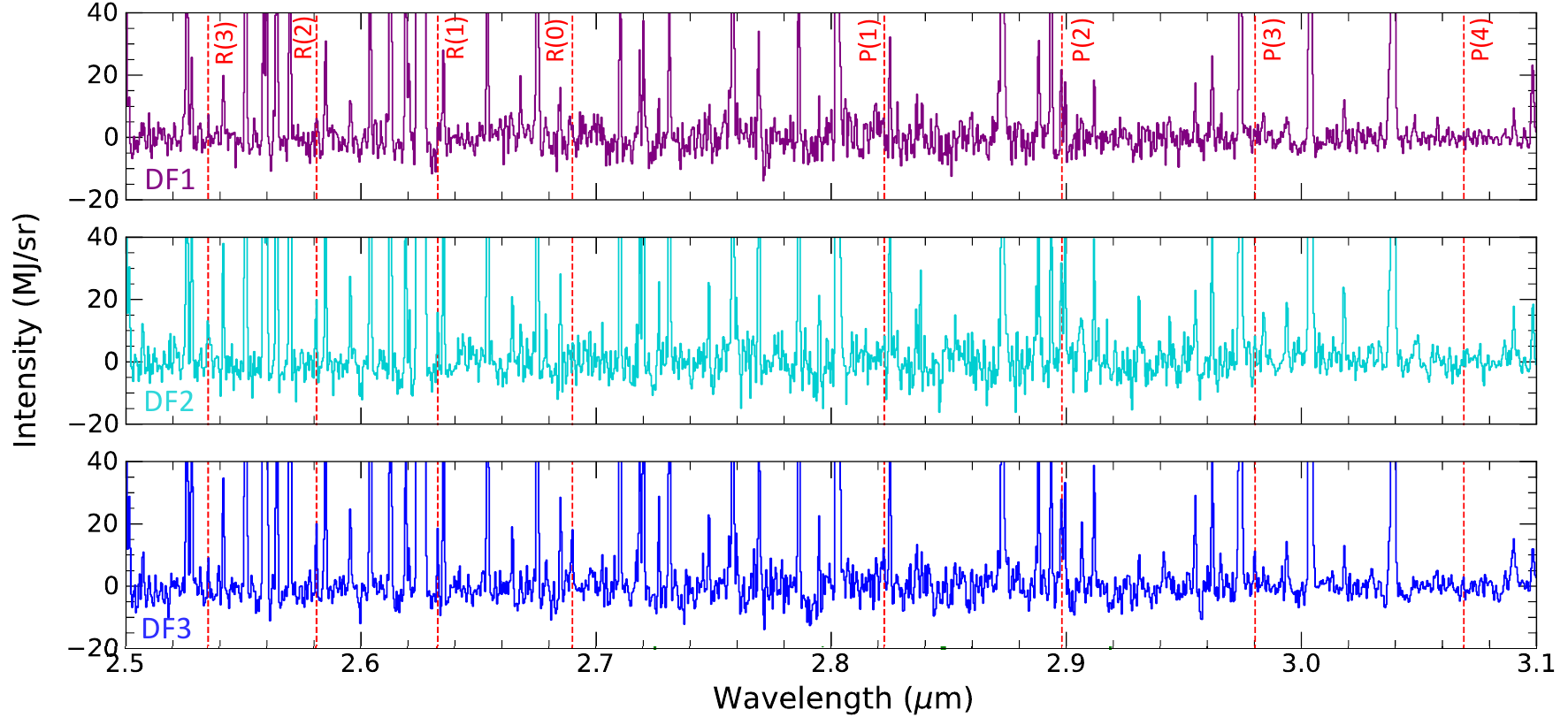}
    \caption{NIRSpec continuum subtracted observed spectrum. The red vertical dashed lines are the position of the first rovibrational $v = 1 \rightarrow 0$ $\mathrm{HD}$ lines. From top to bottom, this figure shows the spectrum of the dissociation fronts of the Orion Bar: DF1, DF2, DF3. The apertures taken to average these spectra are from \cite{2024A&A...685A..74P} for the dissociation fronts. These spectra are not corrected for extinction.}
    \label{fig:obs_spec_HD}
\end{figure*}

We used the observations of the NIRSpec integral field unit (IFU) mode  from the Early Release Science (ERS) program PDRs4All\footnote{\url{https://pdrs4all.org/}, DOI: 10.17909/pg4c-1737}: Radiative feedback from massive stars \citep[ID1288, PIs: Berné, Habart, Peeters,][]{ERS_2022}. NIRSpec observations cover a $9 \times 1$ mosaic (NIRspec's field of view being 3" $\times$ 3") centered on $\alpha_{\rm J2000} = 05^{\rm h}35^{\rm min}20.4749^{\rm s}$, $\delta_{\rm J2000} = -05^\circ 25$'$10.45$". We use the full spectro-imaging cubes\footnote{Available on MAST PDRs4All website: \url{https://doi.org/10.17909/wqwy-p406}}. The data were reduced using the JWST science pipeline (version 1.16.0) and the context jwst\_1295.pmap of the Calibration References Data System (CRDS) (see \cite{2024A&A...685A..74P} for observation parameters and data reduction process). The world coordinate system (WCS) of the cubes were corrected by measuring the coordinates of the proplyd d203-506, and then computing the offset between the centroid of this proplyd in synthetic F212N images (NIRSpec data integrated over NIRCam filter F212N) and the manually set coordinate values. The same shift was applied to all three NIRSpec channels. We also use the average spectrum extracted over the apertures given by \cite{2024A&A...685A..74P}, \cite{Chown_2024} and \cite{2024A&A...687A..86V} for the Bar (see Fig. \ref{fig:rgb_OB}). These spectra are in units of $\mathrm{MJy}\,\mathrm{sr}^{-1}$.

Analyzing faint $\mathrm{HD}$ lines, we noticed that updates in  
the data reduction pipeline and reference files have affected the measurements, and in particular, there were previous versions of our reduced data where certain lines were no longer detected (for instance, the 1-0 R(3) line). It is very difficult to precisely assign the origin of this effect observed on the spectra but this is probably linked to the reference files (hence the dark and flat used) which have evolved since the commissioning. The most recent data reduction provides the cleanest spectrum and does not seem affected by these effects. However, further investigation is needed and is out of the scope of this paper.

As the $\mathrm{HD}$ rovibrational emission is detected in the NIR, it can be affected by extinction along the line of sight.
We need to correct  $\mathrm{HD}$ emission for extinction by the matter inside the PDR and the foreground matter in the line of sight\footnote{The extinction by the matter inside the PDR in the line of sight is different from the extinction by the matter inside the PDR towards $\theta^1$ Ori C, which is negligible at the $\mathrm{H}/\mathrm{H_2}$ transition ($A_v < 1$) (see Fig. 14 of \citet{2024A&A...685A..74P} and Fig. 5 of \citet{2024A&A...685A..73H}).}. In order to do that, we use the $R(V)$-parameterized average Milky Way curve by \cite{Gordon_2023}.  \citep{Gordon_2009,Fitzpatrick_2019,Gordon_2021,Decleir_2022}, for an $R(V) = 5.5$ following \cite{2024A&A...685A..74P} and \cite{2024A&A...687A..86V}. In the third dissociation front (DF3), the extinction in the line of sight is estimated to be around $A_\mathrm{V} = 3.4$, as derived by summing the extinction determined for the foreground ($A_\mathrm{V} = 1.4$) and that intrinsic to the atomic PDR ($A_\mathrm{V} = 2$). In DF2, using the same extinction curve, the extinction is estimated to be around $A_\mathrm{V} = 5.87$. One should note that the extinction curve defined in \citet{Gordon_2023} is not sufficient to explain the observations in the Orion Bar (Meshaka et al in prep). However, the relevant wavelength range of detected $\mathrm{HD}$ lines is very narrow (2.5 - 3.0 $\mu$m), leading to a negligible impact on the results of this study compared to other sources of uncertainties.

\subsection{Detection and spatial morphology}

$\mathrm{HD}$ rovibrational lines are detected in several regions of the Orion Bar, mainly in the three dissociation fronts. Fig. \ref{fig:obs_spec_HD} displays the spectra where $\mathrm{HD}$ lines are detected. In total, we detect $7$ rovibrational lines $v=1 \rightarrow 0$ in DF2 and DF3, from R(0)--R(3) and P(1)--P(3). Fewer lines are detected in the DF1 (only R(1)--R(3)), probably due to the higher extinction toward this front ($A_\mathrm{V} \sim 10$), a result of the terrace-field like structure \citep{2024A&A...685A..73H,2024A&A...685A..74P}.
To derive the absolute intensities of $\mathrm{HD}$ lines, we fitted the observed lines with a Gaussian coupled with a linear function to take into account the continuum and then integrated the Gaussian function over the wavelengths. When necessary, a fit of two gaussians was made for $\mathrm{HD}$ lines which were blended (e.g., 1-0 P(1),1-0 R(3)). The upper limit of intensities are calculated by considering two times the noise level on the continuum. Column densities in the upper level of each transition are calculated as follow:

\begin{equation}
    N_{\rm up} = 4 \pi \frac{I}{A_{ij}} \frac{\lambda}{hc}
\end{equation}

where $I$ is the integrated intensity\footnote{Also called integrated surface brightness.} of the line (erg cm$^{-2}$ s$^{-1}$ sr$^{-1}$), 
$\lambda$ the wavelength of the transition ($\mu$m), $A_{ij}$ the Einstein coefficient of the transition ($\mathrm{s}^{-1}$), $h$ the planck constant and $c$ the velocity of light. Column densities from different lines coming from the same upper level are combined according to the expressions of Appendix \ref{Error_bars}. 
As the lines in DF1
are very faint, we exclude them for the rest of this paper.
Integrated intensities of the lines and column densities of the levels observed in DF3 and DF2 are presented in Table \ref{tab:tab_mesure}.

\begin{table*}
    \centering
        \caption{Integrated intensities and upper level column densities of $\mathrm{HD}$ observed rovibrational lines in DF2 and DF3.}
        \begin{subtable}[t]{\textwidth}
        \centering
        \caption{DF3}
    \begin{tabular}{ccccccc}
    \hline \hline
     \multirow{2}{*}{Line} & $\lambda$ & $E_{\rm up}$   & $I$   & $I_{\rm corr}$\tablefootmark{a} & $N_{\rm up}$  & $N_{{\rm up}_{\rm corr}}$\tablefootmark{a}\tablefootmark{b} \\
      & ($\mu$m) & (K) &  ($\times 10^{-6}$ erg cm$^{-2}$ s$^{-1}$ sr$^{-1}$)  &  ($\times 10^{-6}$ erg cm$^{-2}$ s$^{-1}$ sr$^{-1}$) &  ($\times 10^{12}$ cm$^{-2}$)  &  ($\times 10^{12}$ cm$^{-2}$)   \\
      \hline
      \multicolumn{7}{c}{$v=1$}\\
      \hline
 1--0 P(1) &  2.823&5225.9 &5.3 $\pm$ 4.0 & 7.5 $\pm$ 5.7 & 3.1 $\pm$ 2.3 & 4.4 $\pm$ 3.3  \\
 1--0 P(2)& 2.898 &5348.7  & < 10.9 &	< 15.3  & < 12.9 &  < 18.0  \\
 1--0 R(0) & 2.690& 5348.7 & 7.4 $\pm$ 	1.0 &   	10.8 $\pm$ 1.5 & 7.5 $\pm$ 1.0 & 10.9 $\pm$ 1.4\\
 1--0 P(3)& 2.980&  5593.5 & 4.3 $\pm$ 1.1 &  	5.9 $\pm$ 1.5 & 7.9 $\pm$ 2.0 & 11.0 $\pm$ 2.8   \\
1--0 R(1) & 2.633 & 5593.5 & 9.6 $\pm$ 3.1 & 	14.2 $\pm$ 4.5 & 6.5 $\pm$ 2.0 & 9.6 $\pm$ 3.1   \\
1--0 P(4)& 3.069 & 5958.6 & < 1.6 &	< 2.2 &  < 4.6 & < 6.4 \\
1--0 R(2) & 2.581 & 5958.6 & 9.6 $\pm$ 2.5 &  	14.3 $\pm$ 3.7 & 4.9 $\pm$ 1.3 & 7.4 $\pm$ 1.9  \\ 
 1--0 R(3) & 2.535 & 6441.1 & 5.4 $\pm$	1.2 & 	8.1 $\pm$	1.8 & 2.2 $\pm$ 0.5 & 3.4 $\pm$ 0.8 \\
\hline
\multicolumn{7}{c}{$v=2$}\\
\hline
2--1 R(0)& 2.828 & 10313.6 & < 4.4	& < 6.2 &  < 3.1	&	< 4.4 \\
\hline
    
    \end{tabular}
    \end{subtable}
           \begin{subtable}[t]{\textwidth}
        \centering
        \caption{DF2}
    \begin{tabular}{ccccccc}
    \hline \hline
     \multirow{2}{*}{Line} & $\lambda$ & $E_{\rm up}$   & $I$   & $I_{\rm corr}$\tablefootmark{a} & $N_{\rm up}$  & $N_{{\rm up}_{\rm corr}}$\tablefootmark{a}\tablefootmark{b} \\
      & ($\mu$m) & (K) &  ($\times 10^{-6}$ erg cm$^{-2}$ s$^{-1}$ sr$^{-1}$)  &  ($\times 10^{-6}$ erg cm$^{-2}$ s$^{-1}$ sr$^{-1}$) &  ($\times 10^{12}$ cm$^{-2}$)  &  ($\times 10^{12}$ cm$^{-2}$)   \\
      \hline
      \multicolumn{7}{c}{$v=1$}\\
      \hline
 1--0 P(1) & 2.823&5225.9 & < 16.8 & < 30.8	& < 9.7 & < 17.7 \\
 1--0 P(2)& 2.898 &5348.7  & < 11.3 	& < 20.3 & < 13.4 & < 24.0 \\
 1--0 R(0) & 2.690& 5348.7 & < 3.7 &	< 7.1 & < 3.7 & < 7.2\\
 1--0 P(3)& 2.980&  5593.5 & 4.5 $\pm$ 1.4 &  	7.9 $\pm$ 2.5 & 8.4 $\pm$ 2.7 & 14.8 $\pm$ 4.7  \\
1--0 R(1) & 2.633 & 5593.5 & 7.8 $\pm$ 2.2 &	15.3 $\pm$ 4.3 & 5.3 $\pm$ 1.5 & 10.3 $\pm$ 2.9  \\
1--0 P(4)& 3.069 & 5958.6 & < 2.0& < 3.4 & < 5.8 & < 9.9 \\
1--0 R(2) & 2.581 & 5958.6 & 9.6 $\pm$ 	2.5 &  	19.3 $\pm$ 5.0 &  5.0 $\pm$ 1.3 &  9.9 $\pm$ 2.7 \\ 
 1--0 R(3) & 2.535    & 6441.1 & 7.4 $\pm$ 2.8 & 	15.1 $\pm$ 5.6 & 3.1 $\pm$ 1.2 & 6.3 $\pm$ 2.4\\
\hline
\multicolumn{7}{c}{$v=2$}\\
\hline
2--1 R(0)& 2.828 & 10313.6 & < 4.1 & < 7.5 & < 2.9 & < 5.3 \\
\hline
    
    \end{tabular}
    \end{subtable}

    \label{tab:tab_mesure}
    \tablefoot{The uncertainties do not take into account calibration effects nor the error on the choice of extinction curve but the latter is expected to be negligible compared to the measurement uncertainty. \tablefoottext{a}{Values corrected from extinction using the extinction curve from \citet{Gordon_2023} for an $R_V = 5.5$ and $A_\mathrm{V} = 3.4$ in DF3 and $A_\mathrm{V}  = 5.87$ in DF2.}\tablefoottext{b}{In order to trace the excitation diagram, column densities from different lines coming from the same upper level are combined according to the expressions of Appendix\,\ref{Error_bars}.}}
\end{table*}

\begin{figure}
    \centering
    \includegraphics[width=\linewidth]{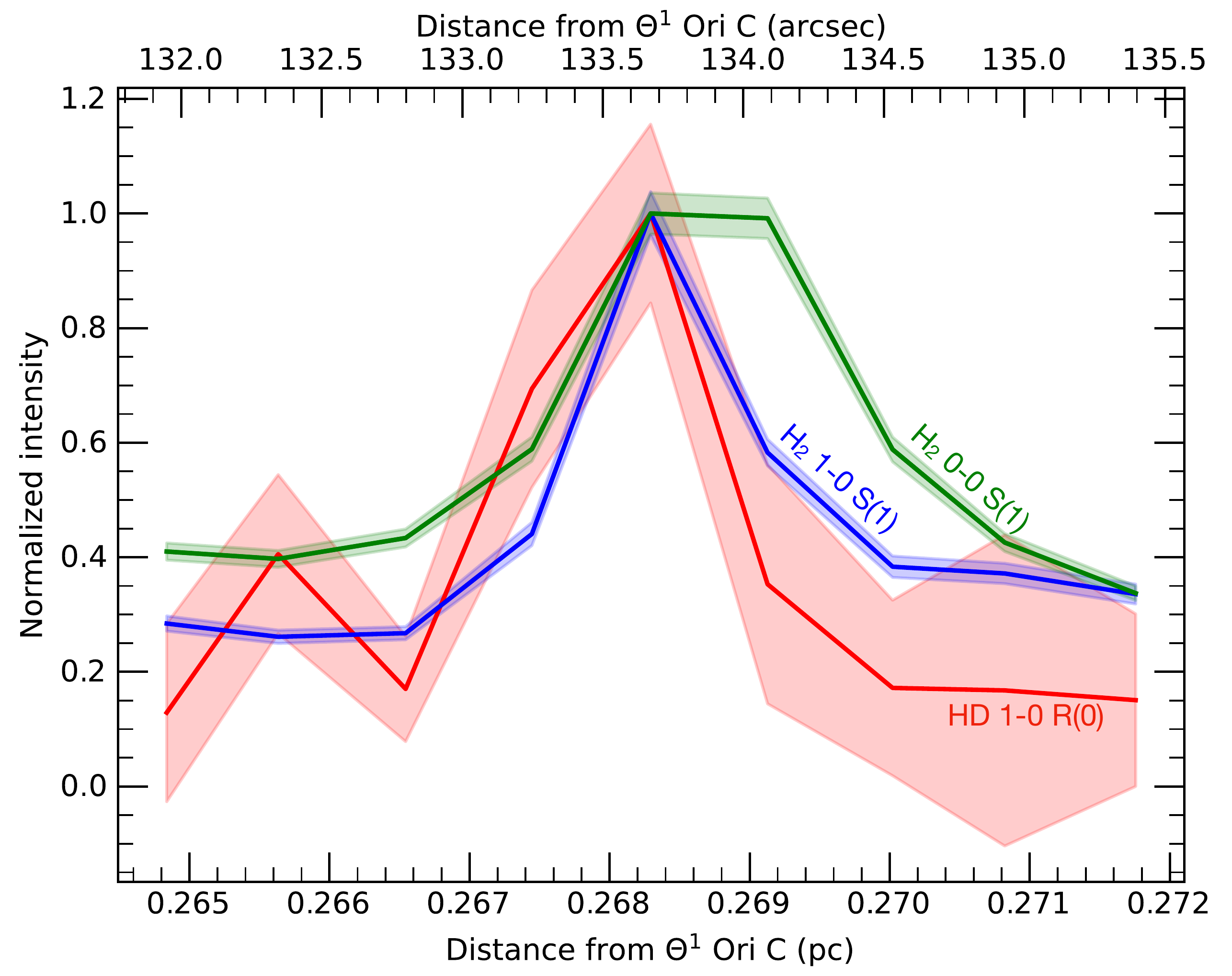}
    \caption{Normalized integrated intensity profiles of the $\mathrm{HD} \, 1-0$ R(0) line, $\mathrm{H}_2 \, 0-0$ S(1) line and $\mathrm{H}_2 \, 1-0$ S(1) line across the third dissociation front DF3 as a function of the projected distance to the star $\Theta^1$ Ori C. The distance to the Orion Bar used to calculate the distance to the star in arcsec is $d=414$ pc \citep{Menten_2007}. Each point corresponds to the intensity averaged on apertures with width of 2.5" and height of 0.4" with an angle of 38$\degree$ (to match DF3 orientation) to increase the S/N.
    }
    \label{fig:cut_hd_obs}
\end{figure}

Fig. \ref{fig:cut_hd_obs} displays the normalized integrated intensity profiles of $\mathrm{HD}$ and $\mathrm{H}_2$ lines around DF3.
It shows that $\mathrm{HD}$ and $\mathrm{H}_2$ rovibrational emission coincide, slightly closer to the edge of the cloud than the pure rotational lines of H$_2$.

\subsection{Excitation}
\label{sect:obs_excit}
\begin{figure}
    \centering
    \includegraphics[width=\linewidth]{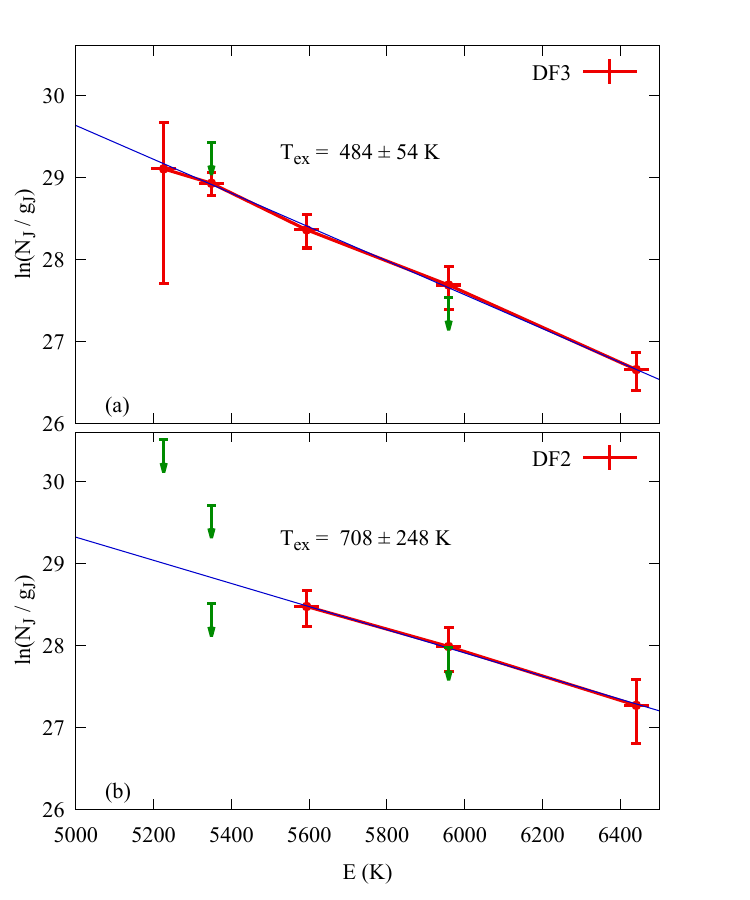}
    \caption{Excitation diagram of $\mathrm{HD}$ rovibrational lines in the DF3 (a) and the DF2 (b). The arrows correspond to the upper limits of $\mathrm{HD}$ lines which can not be properly measured. The solid lines corresponds to the fit to the distribution. The y-axis scales are identical to allow a visual comparison of $T_{\rm ex}$.}
    \label{fig:diag_rot}
\end{figure}

Fig. \ref{fig:diag_rot} displays the rotational diagrams of detected rovibrational $\mathrm{HD}$ lines in DF3 and DF2. In DF3, we can properly derive the column densities of five levels, which
allows to derive an excitation temperature of $T_{\rm ex} \simeq 484 \pm 54\,\mathrm{K}$ following
\begin{equation}
\ln\left(\frac{N_{\rm up}}{g_{\rm up}}\right)= \ln\left(\frac{N}{Q(T_{\rm ex})}\right)-\frac{E_{\rm up}}{k_{\rm B} T_{\rm ex}},
\label{eq:excit_diagram}
\end{equation}

In DF2, fewer lines can be adequately measured, hence the excitation temperature is badly constrained. However, the range of derived values of $T_{\rm ex} \simeq 708 \pm 248\, \mathrm{K}$ encompasses the value derived in DF3. 

Here, the uncertainty is only calculated based on the measurement of $\mathrm{HD}$ line integrated intensities. The detected lines are very weak and so their integrated intensities have large error bars, which is the main cause of uncertainties. However, other factors could affect the estimation of the excitation temperature. First, the correction for extinction could affect the line ratios of $\mathrm{HD}$ and thus the derived excitation temperature. Using Eq. \ref{eq:dT} of Appendix \ref{appendix:uncertainty}, the relative uncertainty on the temperature induced by the limited knowledge of the extinction is at most a few percents, due to the narrow wavelength range. The second factor which can greatly affect the excitation temperature is the uncertainty on integrated intensities due to data reduction effects, as discussed in Sect. \ref{sec:data_reduc}. These effects seem to reduce the intensity of detected lines which lead to large relative uncertainties for weak lines. As we are only detecting a few levels, the absolute intensities of each transition have a great impact on the derived excitation temperature. For instance, if the intensity of 1-0 R(3) is underestimated due to data reduction effects, the derived excitation temperature will be higher. 

Fig. \ref{fig:diag_rot} shows that the five weighted column densities are very well aligned, pointing to the conclusion that the excitation temperature is most likely around $500\, \mathrm{K}$. However, due to the small number of lines used to derive the temperature and the impact of data reduction discussed above, we cannot  
firmly ignore the possibility of a slighly higher or colder excitation temperature. Using the spectra of DF2 and DF3, where we expect the same $\mathrm{HD}$ excitation, the excitation temperature seems to vary between $T_{\rm ex} \sim 480 - 710\,\mathrm{K}$. To be conservative, we will also use this range of values to compare with PDR models.

\section{Comparison between models and observations}\label{Sec:Fit_Obs}

\subsection{Comparison with the fiducial model}

In the previous sections, we presented the observations of rovibrational lines of $\mathrm{HD}$ and discussed their uncertainties. We also presented the main characteristics of a fiducial model and the prediction of $\mathrm{HD}$ abundance and excitation. In this section, we discuss to what extent this kind of model can reproduce the observations and if $\mathrm{HD}$ can be used to constrain physical parameters of PDRs. 
\begin{figure*}
\centering\includegraphics[width=1\linewidth]{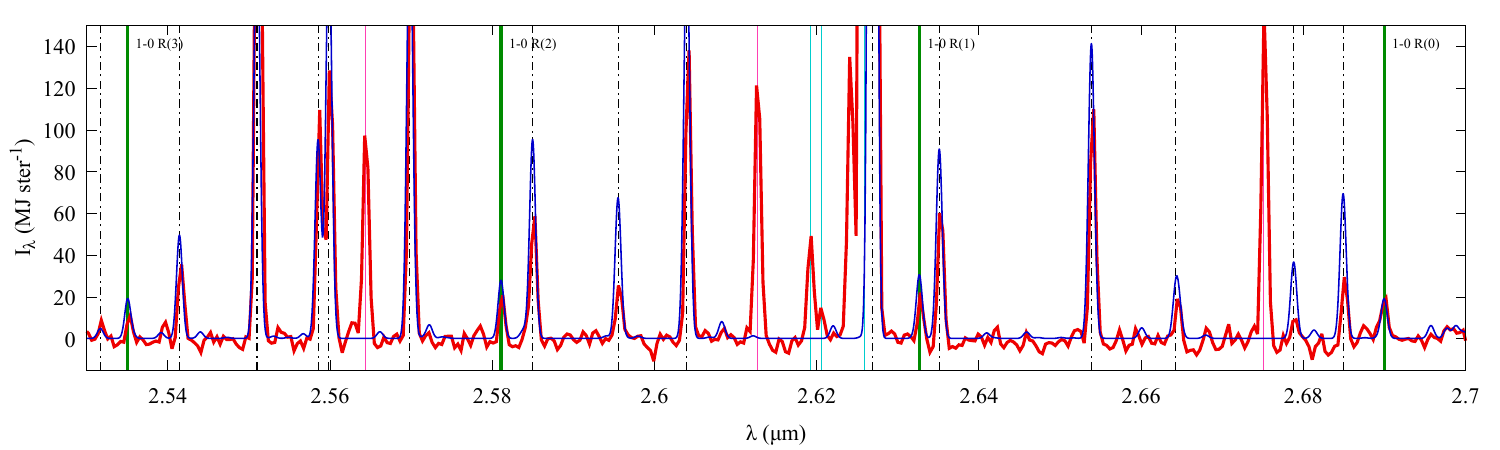}
\caption{Comparison of observational (red) and theoretical (blue) spectra. Dash-dotted vertical lines indicate lines of $\mathrm{H}_2$. Cyan and pink vertical lines indicate recombination lines of $\mathrm{H}$ and $\mathrm{He}$ respectively. Green vertical lines show the location of $\mathrm{HD}$ transitions. The theoretical spectrum is computed for an angle of view of $70^{\circ}$.}
\label{Fig:OBHD_Match_HD}
\end{figure*}

Overall, the predictions of the fiducial model
described in Sect. \ref{Sec:model_Results} are compatible with representative features
of $\mathrm{HD}$ observations, such as the location of the $\mathrm{HD}$ emission and its excitation. Indeed, the model establishes that formation of $\mathrm{HD}$ occurs as soon as some $\mathrm{H_2}$ is available, in zone $II$, and that  deuterium becomes fully molecular slightly before hydrogen as shown in Fig. \ref{Fig:OBHD_zones_P5e7_d5}.  That zone corresponds to the maximum of vibrational excitation favoured by efficient ultraviolet pumping followed by radiative cascades into excited vibrational levels. 
The model can also account for the excitation of $\mathrm{HD}$ in this region. Indeed, the model predicts an excitation temperature $T_{{\rm ex}_{\rm model}} \sim 600 \,\mathrm{K}$, only marginally higher than what is observed $T_{{\rm ex}_{\rm model}} \sim 500 \,\mathrm{K}$.

\begin{figure}
\centering\includegraphics[width=1\columnwidth]{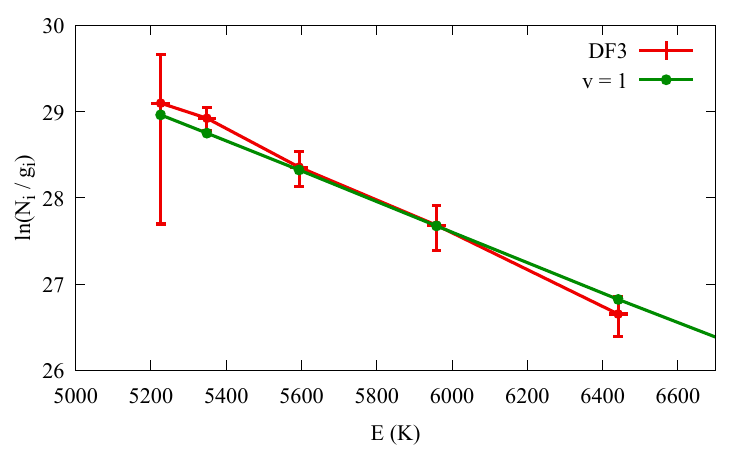}
\caption{Comparison of observational (in red) and theoretical (in green) excitation diagrams for $v=1$. Theoretical column densities have been
computed
for an angle of view of $70^{\circ}$.}
\label{Fig:OBHD_Texcit2_P5e7_d5_Obs_zoom}
\end{figure}

Fig. \ref{Fig:OBHD_Match_HD} shows a comparison of the modeled and observed spectrum for $P = 5\times 10^7\,\mathrm{K}\,\mathrm{cm}^{-3}$ and $d_* = 0.5\,\mathrm{pc}$, with line identification. The theoretical spectrum is computed at an angle of $70^{\circ}$ with respect to the normal of the cloud to account for the fact that the various dissociation fronts are seen almost edge on (see Fig. \ref{fig:rgb_OB}). This angle is chosen to best fit the observed column density. Fig. \ref{Fig:OBHD_Texcit2_P5e7_d5_Obs_zoom} shows the observational and theoretical excitation diagrams for the detected levels of $v=1$. Theoretical values have been scaled by  a factor $1/\cos \left( 70^{\circ} \right)$ here. Given the large sources of uncertainties (See Sect. \ref{Sec:Observations}) and the fact that the front is certainly not straight, the agreement is satisfactory.

\subsection{$\mathrm{HD}$ as a tracer of physical conditions in the Orion Bar}

The underlying question is whether $\mathrm{HD}$ rovibrational emission  
can be used as
a valuable tracer of physical conditions in PDRs. In order to find the constraints that $\mathrm{HD}$  
can provide, we explore a range of physical parameters. With respect to Sect. \ref{Sec:model_Results}, we keep all parameters fixed, except the pressure $P$ and the intensity of the UV field $G_0$. As we focus on constraining the physical conditions in the PDR, we can exclude, as fitted parameters, $A_V$ which is not relevant for $\mathrm{HD}$ lines emitted in a surface layer and $\zeta$ as cosmic rays are not the main energy input in this surface layer. In addition, we also exclude the extinction properties which are related to the material composition. In order to constrain a best-fit model for the Orion Bar, we consider a few pairs of parameter values, with a thermal pressure from $P = 3\, 10^{7}$ to $10^{8}\,\mathrm{K}\,\mathrm{cm}^{-3}$ and $G_0$ from $\sim 0.7\times 10^4$ to $\sim 4\times 10^4$ (by varying the distance to the star $d_*=0.3-0.7$ pc).

To compare with observations, we select two quantitative constraints from the data, which are:
\begin{itemize}
    \item The distance $D$ between the ionization front and the rise of $\mathrm{H_2}$ emission. We adopt a value of $D = 3.8 \pm 1.0 \, 10^{-2} \,\mathrm{pc}$ as discussed in \citet{2024A&A...685A..73H}. In the model, this is $d_I$, the size of zone $I$.
    
    \item The $\mathrm{HD}$ excitation temperature derived from $v = 1$. In Sect. \ref{Sec:Observations}, we get $T_{\rm ex} = 484 \pm 54\,\mathrm{K}$. As discussed in Sect. \ref{sect:obs_excit}, due to data reduction effects on line intensities and the uncertainties on extinction correction, we cannot firmly rule out the possibility of a slighly higher or colder excitation temperature.
    To be conservative, we will also consider a range of value from  $T_{\rm ex} = 480 - 710\,\mathrm{K}$ to compare to models.
\end{itemize}

\begin{figure}
\centering\includegraphics[width=1\columnwidth]{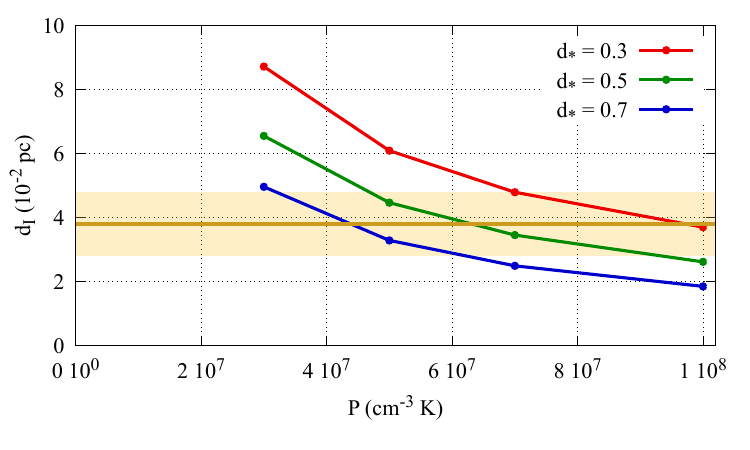}
\caption{Size of zone $I$ for PDR models at different thermal pressure and UV field intensity (scaled by the distance to the star).}
\label{Fig:OBHD_Size_I}
\end{figure}

\begin{figure}
\centering\includegraphics[width=1\columnwidth]{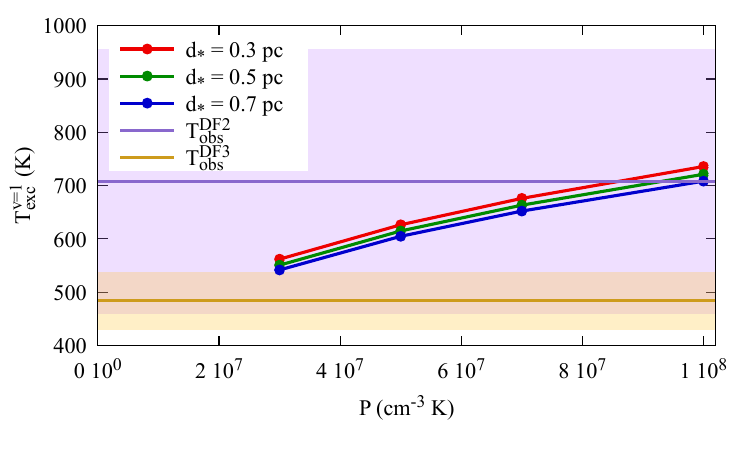}
\caption{$v = 1$ excitation temperature for various values of $P$ and $d_*$.
The observational range of excitation temperatures is displayed as
a colored band: yellow for DF3 and purple for DF2.}
\label{Fig:OBHD_Texc}
\end{figure}

Fig. \ref{Fig:OBHD_Size_I} shows the computed values of $d_I$, and Fig. \ref{Fig:OBHD_Texc} the computed values of $T_{\rm ex}$. $T_{\rm ex}$ depends only weakly on the radiation field intensity and is mainly a function of $P$. Hence, the present $\mathrm{HD}$ observations do not strongly constrain the intensity of the UV field. The observational constraint on the size of zone $I$ implies that the thermal pressure lies between  $P = (3-9) \times 10^7$ K cm$^{-3}$. The observed excitation temperature in DF3 around $T_{\rm ex} = 484 \pm 54$ K  
sets an upper limit to $P \leqslant 3\,10^7\, \mathrm{cm}^{-3}\, \mathrm{K}$.
However, if we consider the range of temperature $T_{\rm ex} = 480 - 710$ K, models with pressure ranging from $P = (2 - 9)\times 10^7$ K cm$^{-3}$ can reproduce the observed excitation temperature. Considering both limitations, and being conservative with the comparison between modeled and observed excitation temperature, a range of $P = (3-9) \times 10^7$ K cm$^{-3}$, with no strict boundary on the intensity of the UV field $G_0$, best reproduce present $\mathrm{HD}$ observations. 

This result shows that, due to the great uncertainty of $\mathrm{HD}$ lines intensities and hence excitation temperature, current $\mathrm{HD}$ observations do not provide strong constraints on physical parameters of the Orion Bar. However, the derived range of pressures is in agreement with the latest estimation, $P \sim 5\, 10^7\, \mathrm{K}\, \mathrm{cm}^{-3}$, obtained from  $\mathrm{H}_2$ observations with JWST (Meshaka et al. in prep), which emits in the same region.
This new evaluation is lower than the previous value $P \sim 2 \, 10^8\, \mathrm{K}\, \mathrm{cm}^{-3}$, derived in the Orion Bar with other tracers observed mainly through rotationally excited transitions of $\mathrm{H_2}$, $\mathrm{CO}$, $\mathrm{OH}$, $\mathrm{CH^+}$, ...  with Herschel \citep[e.g.,][]{Joblin_2018}. The range of values obtained from different tracers may result from a pressure gradient inside the PDR, as already mentioned in \cite{Joblin_2018}. Indeed, the tracers observed by Herschel rather come from zone $III$ where collisional excitation is most efficient. We emphasize that we do not try to derive the relative $\mathrm{D/H}$ elemental abundance ratio, that would affect the absolute column density values as well. 

\section{Conclusions}\label{Sec:Conclusions}

In this work, we present a detailed chemical modeling of the radiation induced $\mathrm{D}/\mathrm{HD}$ transition including  rovibrationally excited $\mathrm{HD}$ in a plane parallel slab of gas and dust.
 We updated the Meudon PDR code to include the contribution of rovibrational  excitation processes (collisions, UV pumping, state-to-state chemical reactions, ...) 
 thanks to  recently available molecular data. We also carefully reanalyse 
  the observed rovibrational $\mathrm{HD}$ emission lines that were detected by JWST, thanks to its high sensitivity and spatial resolution, towards the Orion Bar \citep{2024A&A...685A..74P}, and derive a comprehensive excitation diagram. A fiducial isobaric model with physical parameters which best reproduces other tracers (such as $\mathrm{H_2}$) satisfactorily compares with the observations.

The main conclusions of this study can be summarized as
follows:

\begin{enumerate}
    \item
    In this dense environment and intense UV radiation field conditions, the deuterium molecular reservoir $\mathrm{HD}$ is closely linked to the formation of molecular gas.

    \item
    Rovibrational emission of $\mathrm{HD}$ is predicted to peak in a zone where the gas is still partially atomic. Alternatively, purely rotational emission of $\mathrm{HD}$ reaches its maximum deeper in the cloud, after the $\mathrm{H}/\mathrm{H}_2$ transition. Hence, these two emissions trace different regions of the PDR. These predictions are in broad agreement with observations of $\mathrm{HD}$ rovibrational lines peaking where vibrationally excited $\mathrm{H}_2$ is bright, close to the dissociation front. The detailed comparison of the involved length scales is beyond the present study.

    \item
    The state-to-state reaction rate coefficients for the $\mathrm{D}+ \mathrm{H_2} \leftrightarrow  \mathrm{H} + \mathrm{HD}$ reactions have been introduced into a PDR model thanks to detailed theoretical computations \citep{2019JChPh.150h4301B,10.1063/5.0017697}. This achievement allows to confirm the role of vibrationally excited $\mathrm{H_2}$ in the formation of vibrationally excited $\mathrm{HD}$ in this strong PDR environment, as previously emphasized for the detection of $\mathrm{CH^+,\, SH^+}$
    \citep{2010ApJ...713..662A,2022A&A...664A.190G}.

    \item 
    The excitation diagram derived from the five observed levels of $\mathrm{HD}$ ($v=1$, $J=0-4$) gives an excitation temperature close to $T_{\rm ex} \sim 500 \, \mathrm{K}$. This temperature is however rather uncertain due to the faintness of the transitions, resulting in low signal to noise, and a limited number of detected levels. Data reduction effects may also affect the measured intensity of the faintest observed lines. A conservative range of observationally derived excitation temperatures is $T_{\rm ex} = 480 - 710\, \mathrm{K}$, based on the hypothesis that DF2 and DF3 have similar excitation conditions.

    \item 
     We find that models with a range of pressure $P = (3-9) \times 10^7 \, \mathrm{K}\,\mathrm{cm}^{-3}$ are compatible with observations. This range of values is in agreement with the latest estimation of the thermal pressure in the Orion Bar derived from the analysis of $\mathrm{H}_2$ emission (Meshaka et al. in prep). This range of pressure is lower than the previously derived value from rotational spectra of $\mathrm{H_2}, \mathrm{CO}$, ... that arise in a different zone of the PDR,  suggesting the occurrence  of a  pressure gradient.

     \item
     The study of $\mathrm{HD}$ is not sufficient to constrain the incident UV field as the excitation temperature of $\mathrm{HD}$ rovibrational lines only depends weakly on this parameter.

    \item
    The models also predict that the excitation temperature of $\mathrm{HD}$ rotational emission is close to the gas temperature at the peak position, whereas the excitation of rovibrational levels is subthermal. This may be explained by the major contribution of collisional excitation for rotational levels whereas rovibrational levels are mostly populated by UV radiative pumping followed by the IR cascade.

\end{enumerate}

This reanalysis of challenging and exciting first observations of several vibrationally excited transitions of $\mathrm{HD}$ provides a unique opportunity to undertake a comprehensive survey of physical processes at work in the molecule formation and excitation. It also allows to review in details the transition from the hot atomic layer to the dense molecular inner region. Deeper observations are necessary to detect $\mathrm{HD}$ lines with better signal to noise ratio and to map their emission in order to precisly constrain the thermal pressure towards the $\mathrm{D}/\mathrm{HD}$ and $\mathrm{H}/\mathrm{H_2}$ transitions.

\begin{acknowledgements}
      This work is based [in part] on observations made with the NASA/ESA/CSA James Webb Space Telescope. The data were obtained from the Mikulski Archive for Space Telescopes at the Space Telescope Science Institute, which is operated by the Association of Universities for Research in Astronomy, Inc., under NASA contract NAS 5-03127 for JWST. These observations are associated with program ERS1288. This work was supported by the Thematic Action “Physique et Chimie du Milieu Interstellaire” (PCMI) of INSU Programme National “Astro”, with contributions from CNRS Physique \& CNRS Chimie, CEA, and CNES. E.P. acknowledges support from the University of Western Ontario, the Canadian Space Agency (CSA, 22JWGO1-16), and the Natural Sciences and Engineering Research Council of Canada.
\end{acknowledgements}

\begin{appendix}

\section{Conversions}\label{Sec:Conversions}

\subsection{$d_*$ to $G_0$}\label{Sub:G_0}

\begin{figure}
\centering\includegraphics[width=1\columnwidth]{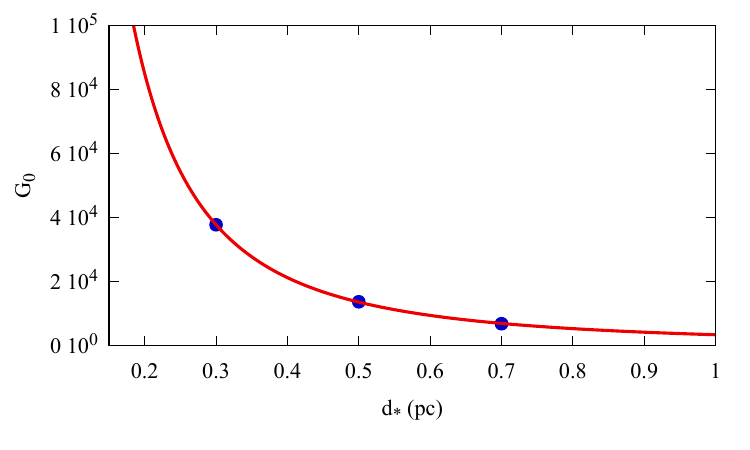}
\caption{Conversion from distance to the star $d_*$ to $G_0$ at the cloud edge.}
\label{Fig:OBHD_G0_d}
\end{figure}

The illuminating star dominates the radiation field intensity at the edge of the cloud, hence $G_0$ scales as $1 / d_*^2$, as shown in Fig. \ref{Fig:OBHD_G0_d}. Values used in this paper are given in Table\,\ref{tab:d-G0.}

\begin{table}
\caption{$d_*$ to $G_0$ conversion.\protect\label{tab:d-G0.}}
\centering%
\begin{tabular}{cccc}
\hline\hline
 $d_*\, (\mathrm{pc})$ & $0.7$ & $0.5$ & $0.3$\tabularnewline
\hline
$G_0$ & $0.7\,10^4$ & $1.4\,10^4$ & $3.8\,10^4$\tabularnewline
\hline
\end{tabular}
\end{table}

\subsection{$d$ to $A_\mathrm{V}$}

\begin{figure}
\centering\includegraphics[width=1\columnwidth]{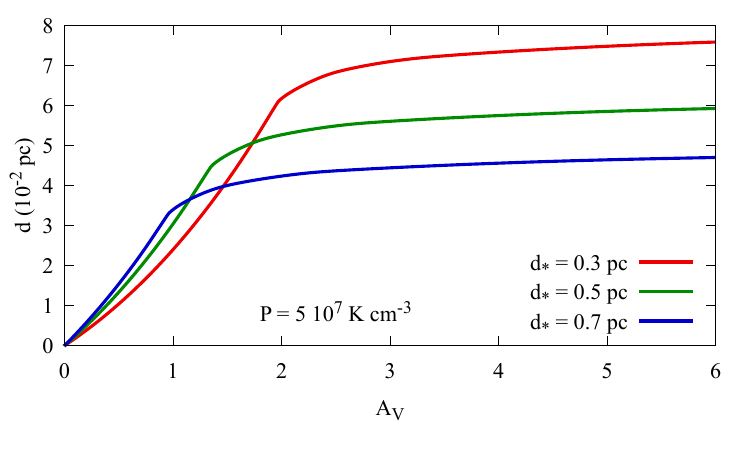}
\caption{$A_\mathrm{V}$ to size conversion.}
\label{Fig:OBHD_AV2d}
\end{figure}

Fig. \ref{Fig:OBHD_AV2d} shows the conversion from $A_\mathrm{V}$ to physical size for different radiation fields.
\begin{equation}
    d = \frac{N_\mathrm{H}}{E_{\mathrm{B-V}}}\, \frac{1}{R_\mathrm{V}}\, \int \frac{dA_\mathrm{V}}{n_\mathrm{H}\left( A_\mathrm{V} \right) }
\end{equation}
where $\frac{N_\mathrm{H}}{E_{\mathrm{B-V}}}$ and $R_\mathrm{V}$ are given in Table\,\ref{tab:Reference-model-parameters.}.

\section{$\mathrm{HD}$ molecular properties}
\label{Sect:HD_details}

We summarize here further molecular properties of $\mathrm{HD}$ that are relevant to our study.

\subsection{Photodissociation}\label{Sub:Photodissociation}

The photodissociation mechanism of interstellar $\mathrm{HD}$ is very similar to $\mathrm{H}_2$: absorption from a low lying rovibrational level of $X$ to an upper electronic excited state is followed by a radiative decay towards the continuum of the $X$ ground electronic state. Hence, for a transition from level $"l"$ (low) in the ground electronic state $X$ to level $"u"$, belonging to the upper electronic state,  the efficiency of dissociation depends on four physical quantities:

\begin{enumerate}
    \item
    The mean radiation field intensity over the line profile $\bar{J}_{ul} = \frac{1}{4\pi}\int I_{ul}\,d\Omega$ (where $I_{ul}$ is the specific intensity of the line (in $\mathrm{erg}\, \mathrm{cm}^{-2}\, \mathrm{s}^{-1}\, \mathrm{sr}^{-1}\, \mathrm{\AA}^{-1}$) at the transition wavelength. This is a function of both the external radiation field intensity and the absorption probability along the line of sight by $\mathrm{HD}$ itself (self-shielding) and by other species, mainly ultraviolet lines of $\mathrm{H}_2$ and continuum absorption by grains and ionisation of light atoms ($\mathrm{C}$ and $\mathrm{S}$). That quantity is obtained from the resolution of the radiative transfer equation as described in \cite{Lepetit:2006}.
    
    \item
    The population $n_l$ of $\mathrm{HD}$ levels at the position of absorption is obtained from the solution to the detailed balance equations \ref{Eq:Steady_state}. It is important to realize that   they may depend on the whole structure of the cloud due to the radiative induced terms.
    
    \item
    The absorption probability that involves the $B_{lu}$ Einstein coefficient is related to the spontaneous emission coefficient $A_{ul}$ through a relation which depends on the units used for the radiation field. For the incident intensity $J$ expressed in $\mathrm{erg}\,\mathrm{s}^{-1}\,\mathrm{cm}^{-2}\,\mathrm{\AA}^{-1}\,\mathrm{sr}^{-1}$, this is:
    \begin{equation}
        B_{lu} = 10^{32}\, \frac{g_u}{g_l}\,\frac{\lambda^5}{2\,h\,c^2}\,A_{ul}\,,
    \end{equation}
    where $\lambda$ is expressed in $\AA$.
    
    \item
    The dissociation probability $p^{dis}_u$, i.e., the fraction of radiative transitions emitted above the dissociation limit of the $X$ $^1\Sigma_g^+$ electronic ground state of $\mathrm{HD}$, that is a property of the upper rovibronic level of the predissociative transition.
\end{enumerate}

\begin{figure}
\centering\includegraphics[width=1\columnwidth]{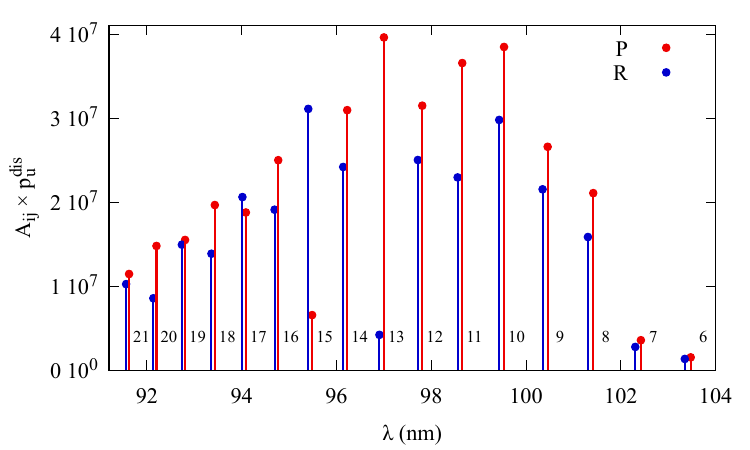}
\caption{Dissociation efficiency factor of $\mathrm{HD}$ lines for transitions starting from $v=0,\,J=2$. Vertical bars are for the $P$ and $R$ lines. The attached number is $v'$.}
\label{Fig:HD_Dis_Weights}
\end{figure}

Then, the dissociation rate, expressed in $\mathrm{cm}^{-3}\, \mathrm{s}^{-1}$ , involving a transition $l - u$, at a specific wavelength $\lambda$, is given by:
\begin{equation}
    k_l = n_l\, \sum_u \, \bar{J}_{ul}\, B_{lu}\, p^{dis}_u = n_l \,  D_l
\end{equation}
 and $\bar{J}_{ul}$ the mean radiation field intensity over the line profile.
The other two parameters are intrinsic properties of the molecule, and they can be evaluated before hand. This is the product $B_{lu}\times p^{dis}_u$, or
alternatively\footnote{Note that the variations of $\lambda$ over the range of absorption wavelengths is small.}
$A_{ul} \times p^{dis}_u$, (hereafter the "dissociation efficiency factor"). Fig. \ref{Fig:HD_Dis_Weights} shows this parameter for the strongest dissociating lines starting from $X$, $v=0,\,J=2$, which is 
significantly populated in zone I (see Figure \ref{Fig:OBHD_v012_P5e7_d5}), and towards levels of $B$ (Lyman system). $A_{ul}$ decreases with $v'$ (upper level vibrational quantum number), in compliance with the Franck-Condon principle, but $p^{dis}_u$ is negligible for small $v'$ values and increases sharply with $v'$. As a result, the most efficient dissociating lines in the  Lyman band system  are $P$ and $R$ transitions involving $v' \sim 8-20$.

\subsection{
 $\mathrm{D} + \mathrm{H}_2  \texorpdfstring{\rightarrow}{→} \mathrm{H} + \mathrm{HD}$
 and $\mathrm{HD} + \mathrm{H}  \texorpdfstring{\rightarrow}{→} \mathrm{H} + \mathrm{HD}$}
\label{Sub:neutral}

The title reactions involving the major neutral hydrogen and deuterium species have a barrier of respectively $E_2 = 3820\,\mathrm{K}$ and $E_5 = 4240\,\mathrm{K}$. 
Their contribution to formation/destruction of $\mathrm{HD}$ depends on whether sufficient energy in the molecular reactant is available to overcome this barrier. \citet{2019JChPh.150h4301B} and \citet{10.1063/5.0017697} have studied these two reactions which lead to the formation / destruction contributions shown in Fig. \ref{Fig:Rxn_FD_HD_d6} in Sect. \ref{Sub:HD_Abundance}  in a state-to-state quasi-classical trajectory approach.

We test these results against two other procedures that are found in the literature, in the absence of such a knowledge, when the reaction rate coefficient is given by an Arrhenius-type formula:
    \begin{equation}
        k_i = \alpha_i\, \left( \frac{T}{300} \right)^{\beta_i} \exp \left(- \frac{E_i}{T} \right),
    \end{equation}

\begin{enumerate}[(1)]
    \item
    No internal energy is available in the molecular reactant and only the kinetic form of energy  (thermal) allows the reaction to occur. This is the approximation usually found in chemical models, where reaction rate coefficients are computed using the previous equation. The values of $\alpha_i$ and $\beta_i$ are found in chemical kinetics databases such as KIDA and UMIST\footnote{\href{https://kida.astrochem-tools.org}{https://kida.astrochem-tools.org} and \href{https://umistdatabase.uk}{https://umistdatabase.uk}}.
    \cite{2011ApJ...737...44G} give analytic expressions of the reaction rate coefficients of both $\mathrm{D}+\mathrm{H}_{2}\rightarrow\mathrm{HD}+\mathrm{H}$ and $\mathrm{HD}  +\mathrm{H}\rightarrow\mathrm{H _{2}} + \mathrm{D}$ reactions as $7.5\, 10^{-11}\, \exp(-3820/T)$ and $7.5\, 10^{-11}\, \exp(-4240/T)$ respectively. These reactions are thus important mainly in the warm front region of the PDR.

    \item
    Including internal excitation of the molecular reactant as introduced by \citet{2010ApJ...713..662A} for the $\mathrm{C^+} + \mathrm{H_2} \rightarrow \mathrm{CH^+} + \mathrm{H}$ endothermic reaction. Each reactive collision occurs with a $\mathrm{H_2}$ molecule in a $(v,J)$ level  of internal energy $E_{v,J}$. 
    This energy is used to overcome the barrier, and  the reaction rate is assumed to be :
    \begin{equation}
        k_i = \alpha_i\, \left( \frac{T}{300} \right)^{\beta_i} \sum_{v,J} x_{v,J}\, \exp \left(- \frac{ \max \left( 0, E_i-E_{v,J} \right)}{T} \right),
    \end{equation}
    where $x_{v,J}$ is the relative abundance of level $(v,J)$. If $E_{v,J} > E_i$, the reaction rate coefficient becomes $ k_i = \alpha_i\, \left( \frac{T}{300} \right)^{\beta_i} \sum_{v,J} x_{v,J}$.
\end{enumerate}

\begin{figure}
\centering\includegraphics[width=1\columnwidth]{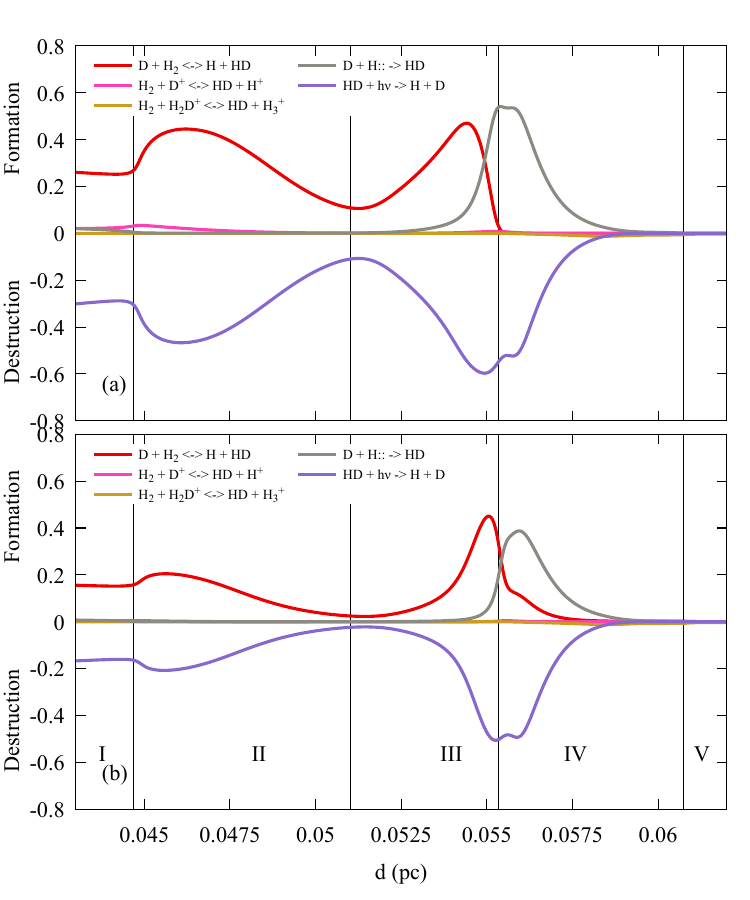}
\caption{Same as Fig. \ref{Fig:Rxn_FD_HD_d6}b for: (a) case (1) and (b): case (2).
}
\label{Fig:Rxn_FD_HD_rb}
\end{figure}

\begin{figure}
\centering\includegraphics[width=1\columnwidth]{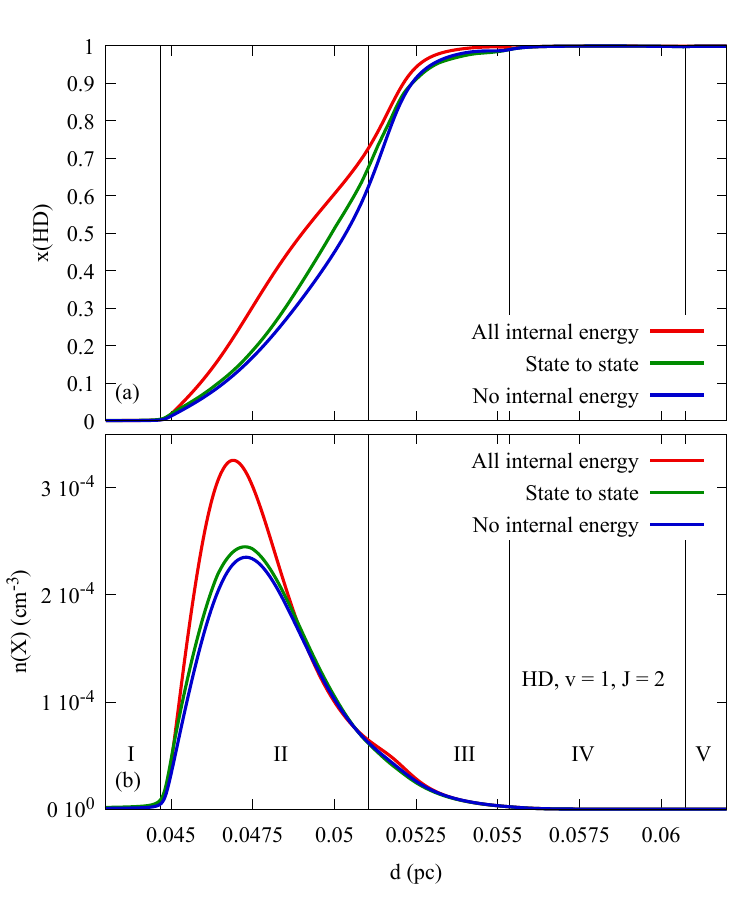}
\caption{(a) Deuterium molecular fraction as a function of distance from the ionization front $d$. Red: case (2), all internal energy used, green: state-to-state chemistry, blue: case (1) analytic expressions from \cite{2011ApJ...737...44G} using no internal energy.
(b) Abundance of level $v=1$ and $J=2$.}
\label{Fig:Cmp_St-St}
\end{figure}

Fig. \ref{Fig:Rxn_FD_HD_rb} shows the modifications to $\mathrm{HD}$ formation / destruction under these two hypotheses as compared to Fig. \ref{Fig:Rxn_FD_HD_d6}b.
Fig. \ref{Fig:Cmp_St-St} shows the fractional abundance of $\mathrm{HD}$ (panel (a)) and the total abundance of one selected vibrational level (panel (b), $v=1,\, J=2$) for the three possible rate coefficients considered here.
Panel (a) shows that internal energy allows $\mathrm{HD}$ to form efficiently closer to the edge of the cloud, thanks to the efficiency of reactions with excited $\mathrm{H}_2$. As a consequence, vibrationally excited levels are more abundant, which leads to higher vibrational lines intensities.

 We notice that the detailed picture of state-to-state reactivity leads to a significant increase of computing time. However, the importance of accounting for internal energy of reactants and products\footnote{This may also involve excited metastable  atomic levels such as $\mathrm{O}\, ^1\mathrm{D}$ or $^1\mathrm{S}$} is recognized in various astrochemical studies 
\citep{2022A&A...664A.190G} but very few detailed results are available up to now.  

\section{Chemical heating}\label{Sec:Chem_Heat}

\begin{figure}
\centering\includegraphics[width=1\columnwidth]{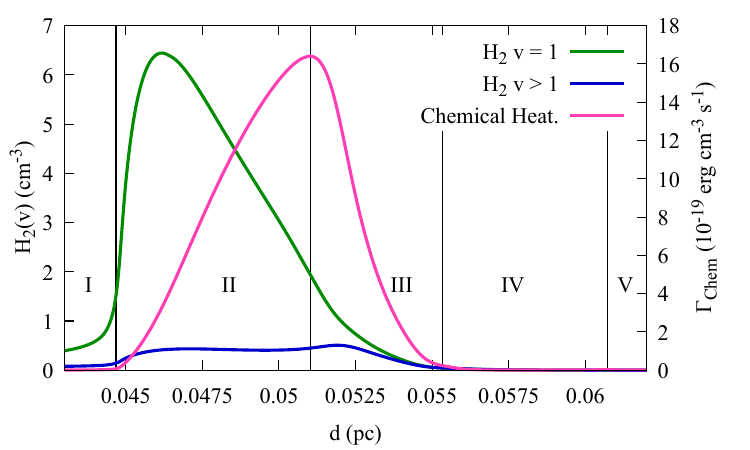}
\caption{Left axis: abundance of vibrationally excited $\mathrm{H}_2$ as a function of distance from the ionization front $d$ in the fiducial model. Right axis: Chemical heating contribution.}
\label{Fig:OBHD_H2v_Lin}
\end{figure}

\begin{figure}
    \centering\includegraphics[width=1\columnwidth]{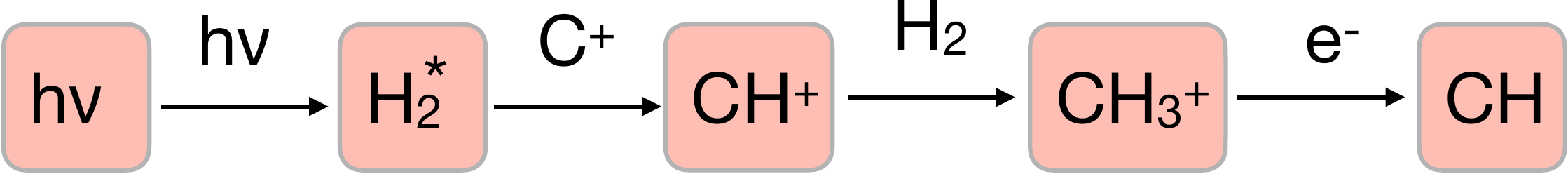}
    \caption{Energy transfer from the ultraviolet external bath to internal heating of the gas by chemistry.}
    \label{Fig:Chem_Heat}
\end{figure}

Chemical heating is hard to evaluate in a low density medium where collisional equilibrium does not dominate and thermal equilibrium is not achieved. Ideally, one should use the full state dependent cross-sections for both reactants and products, and sum over their energy distribution
to compute the balance of the mean kinetic energy.
This is clearly impossible in the present case, and one has to resort to cruder approximations. The exchange of chemical energy  is expressed as the product of the enthalpy variation involved in the reaction by  the number of reactions per unit time and per unit space (i.e., the reaction rate coefficient used to solve the chemical composition of the gas, and which is already known) i.e. $n(A)\, n(B)\, k_{AB}\,  {\Delta H}$ in $\mathrm{cm^{-3} \, \mathrm{erg}\, \mathrm{s}^{-1}}$ for a binary reaction $\mathrm{A + B \rightarrow C + D + \Delta H}$ \citep{atkins2009quanta}

$k_{AB}$ includes any impact of a possible barrier. $\Delta H$ is an intrinsic property of the reaction. From standard text books on thermochemistry (see, e.g. \citet{atkins2009quanta}), it is equal to the thermal energy given to or removed from the gas at constant pressure, including all internal excitation effects, for a medium connected to a heat reservoir.
This is not exactly the case in a low density medium because other physical processes, such as spontaneous radiative transitions, modify the internal distribution of energy of the molecules. However, we have no way to evaluate their possible impact.

The main input energy is provided by the external incident radiation field and the cosmic ray particles.
We do not consider possible mechanical energy due to shocks nor other 
contributions due to the presence of a magnetic field. Cosmic rays dominate deep into the cloud, but ultraviolet photons are the main source of energy in the outer shells. This radiative energy may be converted into internal or kinetic energy through different coupling mechanisms. Photoelectric effect on grains is very efficient to convert radiative energy into gas heating but gas-phase processes can also take place for such a conversion of energy, as illustrated in Fig. \ref{Fig:Chem_Heat}.

\begin{enumerate}[(1)]
\item 
First, external UV photons energy is transferred to $\mathrm{H}_2$ vibrational levels by absorption followed by cascades, with a $90 \%$ probability. As vibrational levels of $\mathrm{H}_2$ are located above $6000\,\mathrm{K}$ (corresponding to $v=1$) above the fundamental level, they may provide a significant energy reservoir if they become populated.

\item 
A part of the internal energy of $\mathrm{H}_2$, excited by UV pumping, is then transferred to chemical potential energy through the  $\mathrm{C}^+ + \mathrm{H_2}$ reaction. 
This endothermic formation reaction of $\mathrm{CH}^+$ requires some energy input to proceed. The internal energy of $\mathrm{H_2}$ has been first suggested by \cite{1974MNRAS.168P..51S} for explaining the presence of ubiquitous $\mathrm{CH^+}$ in diffuse environments. This suggestion has been further quantified and established \citep{Hierl_1997,2010ApJ...713..662A,Naylor_2010,Godard_2013,Nagy_2013}. In the present model, we use a combination of \citet{2013ApJ...766...80Z} and \citet{2014PCCP...1624800H} theoretical computations to build temperature and $\mathrm{H}_2\,(v,J)$ state dependent rate coefficients
for this reaction. 

\item 
$\mathrm{CH^+}$ then reacts with $\mathrm{H_2}$, through  exothermic reactions to form $\mathrm{CH_2^+}$ and $\mathrm{CH_3^+}$. 

\item 
These ions finally recombine with electrons through exothermic reactions, releasing as much as a few $\mathrm{eV}$ in the gas.  

\end{enumerate}

Fig. \ref{Fig:OBHD_H2v_Lin} shows the abundance of vibrationally excited $\mathrm{H}_2$ and the contribution of chemistry to gas heating, as computed from the fiducial model. The later maximum is shifted towards the interior of the cloud because radiative and collisional deexcitation are more efficient than chemistry close to the edge of cloud.The process ceases where no vibrationally excited $\mathrm{H}_2$ remains.

\section{Uncertainties and error bars}

\subsection{Extinction curve induced uncertainties}\label{appendix:uncertainty}

The uncertainty on the extinction curve can be written as an uncertainty on $R_V$ and $A_V$. If we consider that $R_V$ can be in the range $R_V = 3.1 - 5.5$ (from a diffuse medium to a medium with a depletion of nanograins) and $A_V = 3.4 \pm 1.0$ in DF3. To maximize the uncertainty, we consider that the highest extinction possible is when considering an extinction curve with $R_\mathrm{V} = 5.5$ with $A_\mathrm{V} = 4.4$ and the lowest extinction possible  when considering an extinction curve with $R_\mathrm{V} = 3.1$ with $A_\mathrm{V} = 2.4$. Then, the relative uncertainty on the line ratio due to the choice of the slope of the extinction curve can be written as :

\begin{multline}
    \frac{d(I_1/I_2)}{I_1/I_2}_{\rm extinction} = \\
     \frac{e^{-\frac{A_\mathrm{V}+dA_\mathrm{V}}{2.5\log(e)}(\frac{A_{{\lambda}_1}}{A_\mathrm{V}} -  \frac{A_{{\lambda}_2}}{A_\mathrm{V}})\left(R_\mathrm{V}=5.5\right)} - e^{-\frac{A_\mathrm{V}-dA_\mathrm{V}}{2.5\log(e)}(\frac{A_{{\lambda}_1}}{A_\mathrm{V}} -  \frac{A_{{\lambda}_2}}{A_\mathrm{V}})\left(R_\mathrm{V}=3.1\right)}}{e^{-\frac{A_\mathrm{V}}{2.5\log(e)}(\frac{A_{{\lambda}_1}}{A_\mathrm{V}} -  \frac{A_{{\lambda}_2}}{A_\mathrm{V}})\left(R_\mathrm{V}=5.5\right)}}
\end{multline}

The excitation temperature is : \begin{equation}
    T_{\rm ex} = \frac{E_2-E_1}{k_{\rm B}} / \ln \left(\frac{N_1 g_2}{N_2 g_1}\right). 
    \label{eq:Tex}
\end{equation}

Hence, the relative uncertainty on the temperature is : 

\begin{equation}
    \frac{dT}{T}_{\rm extinction} = \frac{k_BT}{E_2-E_1}\frac{d(I_1/I_2)}{I_1/I_2}_{\rm extinction}
    \label{eq:dT}
\end{equation}

Using the 1-0 P(1) ($\lambda_1 = 2.82$ $\mu$m, $E_1 = 5225$ K)  and 1-0 R(3) line ($\lambda_1 = 2.53$ $\mu$m, $E_2 = 6441$ K) and $A_\mathrm{V} \sim 3$, we derive $\frac{d(I_1/I_2)}{I_1/I_2} \sim 0.06$. In DF3, $T_{\rm ex} = 484$ K, thus $\frac{dT}{T}_{\rm extinction} \sim 0.03$.

\subsection{Error bars on mean values and fit parameters}\label{Error_bars}

If two lines are detected coming from the same upper level, the derived column densities are combined to reduce the uncertainties. Each measurement is issued from independent distributions with the same mean. Hence we can combine them under a maximum likelyhood hypothesis. Given two measures $(N_1,\sigma_1)$ and $(N_2,\sigma_2)$, the mean and uncertainty are given by:

\begin{equation}
    N_J = \frac{\frac{N_1}{\sigma_1^2} + \frac{N_2}{\sigma_2^2}}{\frac{1}{\sigma_1^2} + \frac{1}{\sigma_2^2}} \, ;\quad
    \sigma_J = \sqrt{ \frac{1}{\frac{1}{\sigma_1^2} + \frac{1}{\sigma_2^2}}}
\end{equation}

Excitation temperatures obtain from linear fit on the quantities $y_J = \ln{\frac{N_J}{g_J}}$. Given an uncertainty $\sigma_J$ on $N_J$, the uncertainty on $y_J$ is $\frac{\sigma_J}{N_J}$. Note that the factor $g_J$ simplifies from the ratio.

\section{Line intensities}\label{Line_int_fiducial}

In this Appendix, we give a list of $\mathrm{HD}$ line intensities for the fiducial model $P=5\, 10^7\, \mathrm{K\,cm^{-3}}$ and $d_* = 0.5\,\mathrm{pc}$. The choice of lines is limited to $I_{Th} > 10^{-7}\, \mathrm{erg}\, \mathrm{cm}^{-2}\, \mathrm{s}^{-1}\, \mathrm{sr}^{-1}$. 
We note that \citet{Joblin_2018} report a value of $11 \pm 4\, 10^{-5}\, \mathrm{erg}\, \mathrm{cm}^{-2}\, \mathrm{s}^{-1}\, \mathrm{sr}^{-1}$ for the pure rotational $R(1)$ line, detected with Herschel, within a factor of $2$ from our results.
All $Q$ lines and quadrupolar transitions are below this threshold. Most of these lines are too weak to be seen with today's observing facilities, but the full set may be of interest for comparison purposes.

\begin{table}
\caption{$v=0 \rightarrow v=0$ lines intensities.\label{tab:Lines_0_0}}
\centering%
\begin{tabular}{c|cc}
\hline\hline
Line & $\lambda$ & $I$ \tabularnewline
\hline 
$R(0)$ & $112.07$ & $2.61\,(-5)$ \tabularnewline
$R(1)$ & $56.23$ & $6.21\,(-5)$ \tabularnewline
$R(2)$ & $37.70$ & $1.03\,(-4)$ \tabularnewline
$R(3)$ & $28.50$ & $9.62\,(-5)$ \tabularnewline
$R(4)$ & $23.03$ & $4.97\,(-5)$ \tabularnewline
$R(5)$ & $19.43$ & $1.50\,(-5)$ \tabularnewline
$R(6)$ & $16.89$ & $3.06\,(-6)$ \tabularnewline
$R(7)$ & $15.02$ & $6.05\,(-7)$ \tabularnewline
$R(8)$ & $13.60$ & $1.89\,(-7)$ \tabularnewline
\hline 
\end{tabular}
\tablefoot{Line intensities for the fiducial case $P=5\,10^7 \,\mathrm{K\,cm^{-3}}$ and $d_* = 0.5\,\mathrm{pc}$. Wavelengths are in $\mu\mathrm{m}$, $I$ is the integrated intensity in $\mathrm{erg}\, \mathrm{cm}^{-2}\, \mathrm{s}^{-1}\, \mathrm{sr}^{-1}$}
\end{table}

\begin{table}
\caption{$v=1 \rightarrow v=0$ lines intensities.\label{tab:Lines_1_0}}
\centering%
\begin{tabular}{c|cc}
\hline\hline
Line & $\lambda$ & $I$ \tabularnewline
\hline 
$P(1)$ & $2.823$ & $3.99\,(-6)$ \tabularnewline
$P(2)$ & $2.898$ & $4.76\,(-6)$ \tabularnewline
$P(3)$ & $2.980$ & $3.31\,(-6)$ \tabularnewline
$P(4)$ & $3.069$ & $1.57\,(-6)$ \tabularnewline
$P(5)$ & $3.165$ & $5.43\,(-7)$ \tabularnewline
$P(6)$ & $3.268$ & $1.44\,(-7)$ \tabularnewline
\hline
$R(0)$ & $2.690$ & $5.42\,(-6)$ \tabularnewline
$R(1)$ & $2.633$ & $8.80\,(-6)$ \tabularnewline
$R(2)$ & $2.581$ & $8.39\,(-6)$ \tabularnewline
$R(3)$ & $2.535$ & $5.67\,(-6)$ \tabularnewline
$R(4)$ & $2.494$ & $3.01\,(-6)$ \tabularnewline
$R(5)$ & $2.459$ & $1.32\,(-6)$ \tabularnewline
$R(6)$ & $2.428$ & $4.95\,(-7)$ \tabularnewline
$R(7)$ & $2.402$ & $2.40\,(-7)$ \tabularnewline
\hline
\end{tabular}
\tablefoot{Same as Table\,\ref{tab:Lines_0_0}.}
\end{table}

\begin{table}
\caption{$v=2 \rightarrow v=1$ lines intensities.\label{tab:Lines_2_1}}
\centering%
\begin{tabular}{c|cc}
\hline\hline
Line & $\lambda$ & $I$ \tabularnewline
\hline
$P(1)$ & $2.968$ & $1.91\,(-6)$ \tabularnewline
$P(2)$ & $3.048$ & $2.23\,(-6)$ \tabularnewline
$P(3)$ & $3.135$ & $1.53\,(-6)$ \tabularnewline
$P(4)$ & $3.229$ & $7.54\,(-7)$ \tabularnewline
$P(5)$ & $3.331$ & $2.71\,(-7)$ \tabularnewline
\hline
$R(0)$ & $2.828$ & $2.60\,(-6)$ \tabularnewline
$R(1)$ & $2.767$ & $4.25\,(-5)$ \tabularnewline
$R(2)$ & $2.713$ & $4.27\,(-6)$ \tabularnewline
$R(3)$ & $2.665$ & $3.08\,(-6)$ \tabularnewline
$R(4)$ & $2.622$ & $1.78\,(-6)$ \tabularnewline
$R(5)$ & $2.585$ & $7.73\,(-7)$ \tabularnewline
$R(6)$ & $2.553$ & $3.26\,(-7)$ \tabularnewline
\hline
\end{tabular}
\tablefoot{Same as Table\,\ref{tab:Lines_0_0}.}
\end{table}

\begin{table}
\caption{$v=3 \rightarrow v=2$ lines intensities.\label{tab:Lines_3_2}}
\centering%
\begin{tabular}{c|cc}
\hline\hline
Line & $\lambda$ & $I$ \tabularnewline
\hline
$P(1)$ & $3.126$ & $8.53\,(-7)$ \tabularnewline
$P(2)$ & $3.210$ & $9.86\,(-7)$ \tabularnewline
$P(3)$ & $3.303$ & $6.84\,(-7)$ \tabularnewline
$P(4)$ & $3.404$ & $3.42\,(-7)$ \tabularnewline
$P(5)$ & $3.512$ & $1.28\,(-7)$ \tabularnewline
\hline
$R(0)$ & $2.977$ & $1.17\,(-6)$ \tabularnewline
$R(1)$ & $2.914$ & $1.96\,(-6)$ \tabularnewline
$R(2)$ & $2.856$ & $2.04\,(-6)$ \tabularnewline
$R(3)$ & $2.805$ & $1.57\,(-6)$ \tabularnewline
$R(4)$ & $2.761$ & $1.02\,(-6)$ \tabularnewline
$R(5)$ & $2.722$ & $4.66\,(-7)$ \tabularnewline
\hline
\end{tabular}
\tablefoot{Same as Table\,\ref{tab:Lines_0_0}.}
\end{table}

\begin{table}
\caption{$v=2 \rightarrow v=0$ lines intensities.\label{tab:Lines_2_0}}
\centering%
\begin{tabular}{c|cc}
\hline\hline
Line & $\lambda$ & $I$ \tabularnewline
\hline
$P(1)$ & $1.429$ & $1.98\,(-6)$ \tabularnewline
$P(2)$ & $1.449$ & $2.66\,(-6)$ \tabularnewline
$P(3)$ & $1.471$ & $2.16\,(-6)$ \tabularnewline
$P(4)$ & $1.495$ & $1.29\,(-6)$ \tabularnewline
$P(5)$ & $1.520$ & $5.88\,(-7)$ \tabularnewline
$P(6)$ & $1.548$ & $2.20\,(-7)$ \tabularnewline
\hline
$R(0)$ & $1.395$ & $2.12\,(-6)$ \tabularnewline
$R(1)$ & $1.380$ & $3.14\,(-6)$ \tabularnewline
$R(2)$ & $1.369$ & $2.89\,(-6)$ \tabularnewline
$R(3)$ & $1.358$ & $1.92\,(-6)$ \tabularnewline
$R(4)$ & $1.350$ & $1.03\,(-6)$ \tabularnewline
$R(5)$ & $1.343$ & $4.19\,(-7)$ \tabularnewline
$R(6)$ & $1.339$ & $1.67\,(-7)$ \tabularnewline
\hline
\end{tabular}
\tablefoot{Same as Table\,\ref{tab:Lines_0_0}.}
\end{table}

\begin{table}
\caption{$v=3 \rightarrow v=1$ lines intensities.\label{tab:Lines_3_1}}
\centering%
\begin{tabular}{c|cc}
\hline\hline
Line & $\lambda$ & $I$ \tabularnewline
\hline
$P(1)$ & $1.504$ & $1.69\,(-6)$ \tabularnewline
$P(2)$ & $1.525$ & $2.25\,(-6)$ \tabularnewline
$P(3)$ & $1.548$ & $1.85\,(-6)$ \tabularnewline
$P(4)$ & $1.574$ & $1.13\,(-6)$ \tabularnewline
$P(5)$ & $1.601$ & $5.38\,(-7)$ \tabularnewline
$P(6)$ & $1.631$ & $2.21\,(-7)$ \tabularnewline
\hline
$R(0)$ & $1.468$ & $1.83\,(-6)$ \tabularnewline
$R(1)$ & $1.453$ & $2.77\,(-6)$ \tabularnewline
$R(2)$ & $1.440$ & $2.64\,(-6)$ \tabularnewline
$R(3)$ & $1.429$ & $1.87\,(-6)$ \tabularnewline
$R(4)$ & $1.421$ & $1.12\,(-6)$ \tabularnewline
$R(5)$ & $1.414$ & $4.80\,(-7)$ \tabularnewline
\hline
\end{tabular}
\tablefoot{Same as Table\,\ref{tab:Lines_0_0}.}
\end{table}

\begin{table}
\caption{$v=3 \rightarrow v=0$ lines intensities.\label{tab:Lines_3_0}}
\centering%
\begin{tabular}{c|cc}
\hline\hline
Line & $\lambda$ & $I$ \tabularnewline
\hline
$P(1)$ & $0.973$ & $3.99\,(-7)$ \tabularnewline
$P(2)$ & $0.982$ & $5.59\,(-7)$ \tabularnewline
$P(3)$ & $0.993$ & $4.84\,(-7)$ \tabularnewline
$P(4)$ & $1.005$ & $3.13\,(-7)$ \tabularnewline
$P(5)$ & $1.018$ & $1.59\,(-7)$ \tabularnewline
\hline
$R(0)$ & $0.957$ & $4.02\,(-7)$ \tabularnewline
$R(1)$ & $0.951$ & $5.90\,(-7)$ \tabularnewline
$R(2)$ & $0.947$ & $5.47\,(-7)$ \tabularnewline
$R(3)$ & $0.943$ & $3.78\,(-7)$ \tabularnewline
$R(4)$ & $0.940$ & $2.22\,(-7)$ \tabularnewline
\hline
\end{tabular}
\tablefoot{Same as Table\,\ref{tab:Lines_0_0}.}
\end{table}

\end{appendix}

\end{document}